\documentclass[reprint, amsmath,amssymb,aps,prb, superscriptaddress]{revtex4-2}
\usepackage{graphicx}
\usepackage{dcolumn}
\usepackage{bm}

\usepackage{xcolor}
\usepackage{longtable}
\usepackage{colortbl}%
  
\usepackage{booktabs,tabularx}
\usepackage{dcolumn}
\usepackage{hyperref}

\hypersetup{
    colorlinks = true,
    citecolor = {blue},
    linkcolor = {blue},
    urlcolor  = {blue},
}

\usepackage{array}
\newcolumntype{P}[1]{>{\centering\arraybackslash}p{#1}}
\newcolumntype{M}[1]{>{\centering\arraybackslash}m{#1}}

\begin{document}

\

\

\

\ 

\

\ 

\

\

\title{Magnon-phonon interactions in the spinel compound MnSc$_2$Se$_4$}





\author{J. Sourd}
\affiliation{Hochfeld-Magnetlabor Dresden (HLD-EMFL) and Würzburg-Dresden Cluster of Excellence ct.qmat,
Helmholtz-Zentrum Dresden-Rossendorf, 01328 Dresden, Germany}
\author{Y. Skourski}
\affiliation{Hochfeld-Magnetlabor Dresden (HLD-EMFL) and Würzburg-Dresden Cluster of Excellence ct.qmat,
Helmholtz-Zentrum Dresden-Rossendorf, 01328 Dresden, Germany}
\author{L. Prodan}
\affiliation{Experimental Physics V, University of Augsburg, 86135 Augsburg, Germany}
\affiliation{Institute of Applied Physics, MD 2028, Chisinau, Republic of Moldova}
\author{V. Tsurkan}
\affiliation{Experimental Physics V, University of Augsburg, 86135 Augsburg, Germany}
\affiliation{Institute of Applied Physics, MD 2028, Chisinau, Republic of Moldova}
\author{A. Miyata}
\affiliation{Institute for Solid State Physics, The University of Tokyo, Kashiwa, Chiba 277-8581, Japan}
\author{J. Wosnitza}
\affiliation{Hochfeld-Magnetlabor Dresden (HLD-EMFL) and Würzburg-Dresden Cluster of Excellence ct.qmat,
 Helmholtz-Zentrum Dresden-Rossendorf, 01328 Dresden, Germany}
\affiliation{Institut für Festkörper- und Materialphysik, TU Dresden, 01062 Dresden, Germany}
\author{S. Zherlitsyn}
\affiliation{Hochfeld-Magnetlabor Dresden (HLD-EMFL) and Würzburg-Dresden Cluster of Excellence ct.qmat,
Helmholtz-Zentrum Dresden-Rossendorf, 01328 Dresden, Germany}

\date{\today}\begin{abstract}
We investigated the magnetic and magnetoelastic properties of MnSc$_2$Se$_4$ single crystals at low temperature under a magnetic field directed along the crystallographic [111] axis. The magnetization data at low temperature show a linear increase with magnetic field, until saturation is reached above 15 T. In ultrasound, a longitudinal acoustic mode shows a softening in field, which is absent for a transverse acoustic mode. We discuss these results using a microscopic model based on the framework of linear spin-wave theory. The magnetic and magnetoelastic data are qualitatively reproduced by considering magnon-phonon interactions arising from exchange-striction coupling between the crystal lattice and spin-wave fluctuations in the zero-temperature limit.
\end{abstract}

\maketitle

\section{\label{sec:level1}Introduction}

The spinels constitute a large group of transition-metal oxides or chalcogenides, having the general formula $AB_2X_4$. They show many intriguing magnetic and dielectric properties \cite{lacroix2011introduction,tsurkan2021complexity}. Most of these materials are magnetic insulators or semiconductors, and well described by a hard-sphere ionic picture \cite{hill1979systematics}. The anions $X^{2-}$ are generally O$^{2-}$, S$^{2-}$, Se$^{2-}$, or Te$^{2-}$ and form a face-centered cubic lattice, stabilized by interstitial $A$ and $B$ cations which are generally earth-alkaline or $3d$ transition metals.

In cubic spinels of the form $A^{2+}B_2^{3+}X_4^{2-}$, the $A^{2+}$ ions occupy tetrahedral sites and form a diamond sublattice, whereas the $B^{3+}$ ions occupy octahedral sites and form a pyrochlore sublattice. For cations with partially filled $d$ shells, the possibility of different $A$ and $B$ networks together with different crystal-field environments leads to a very rich physics involving the interplay of spin, charge, orbital, and lattice degrees of freedom. The study of spinels, thus, has historically given some precious light about novel types of magnetic order \cite{neel1948proprietes,yafet1952antiferromagnetic}, magnetic frustration \cite{anderson1956ordering}, and spin-lattice couplings \cite{englman1970cooperative}.

Accordingly, spinel compounds show very rich phase diagrams involving temperature, pressure, and magnetic field, revealing some highly exotic states of matter, such as spin loops in ZnCr$_2$O$_4$ \cite{lee2002emergent}, a magnetization plateau in MnCr$_2$S$_4$ \cite{tsurkan2017ultra}, multiferroicity in CdCr$_2$S$_4$ \cite{hemberger2005relaxor}, spin-dimerization in CuIr$_2$S$_4$ \cite{radaelli2002formation}, and an orbital-glass state in FeCr$_2$S$_4$ \cite{fichtl2005orbital}.

The case when only the $A$ site of the spinel lattice is occupied by magnetic ions has gathered a considerable amount of experimental \cite{fritsch2004spin, tristan2005geometric} and theoretical \cite{bergman2007order,lee2008theory} efforts, since the magnetic ions are arranged on a geometrically frustrated diamond lattice. In this respect, the $A$ site spinels of formula MnSc$_2X_4$ ($X=$ S, Se, space group = $Fd\bar{3}m$) are particularly interesting due to the competing antiferromagnetic and ferromagnetic interactions between the magnetic Mn$^{2+}$ ($S=5/2$) ions.
 
MnSc$_2$S$_4$ shows a very rich phase diagram with the presence of helical, incommensurate, modulated and canted orders, together with a skyrmion phase induced by magnetic field \cite{gao2017spiral, gao2020fractional}. With a particularly low Néel temperature, $T_N=2.3$ K, and a Curie-Weiss temperature of $\Theta_{CW} = -22.9$ K, this compound exhibits a substantial frustration factor $f = |\Theta_{CW}|/T_N \approx 10$.

Extending to the MnSc$_2X_4$ system ($X=$ S, Se), the substitution of sulfur by selenide can be understood as a chemically induced pressure effect. The lattice spacing evolves from $a_0=10.6$ \r{A} in MnSc$_2$S$_4$ \cite{fritsch2004spin} to $a_0 = 11.1$ \r{A} for MnSc$_2$Se$_4$ \cite{guratinder2022magnetic}, thus increasing the distances between the magnetic Mn$^{2+}$ ions and decreasing the magnetic exchange energy. Magnetization and neutron-diffraction measurements on powder samples suggest that MnSc$_2$Se$_4$ exhibits magnetic order below $T_N = 2$ K, with a Curie-Weiss temperature $\Theta_{CW} = -18.4$ K, giving also a substantial frustration factor $f = |\Theta_{CW}|/T_N \approx 9$ \cite{guratinder2022magnetic}. A helical and a modulated magnetic order have been proposed from the neutron data, but no skyrmion phase has been observed, so far, in this compound.

In this paper, we explore the role played by spin-strain interactions in the physics of frustrated spin systems. Probing magnetically frustrated systems through the spin-lattice coupling has become a very fruitful method in recent years \cite{ye2006spontaneous,sushkov2005probing}. In particular, measurement of the sound velocity and sound attenuation provide useful information on the static and dynamic properties of the magnetic fluctuations \cite{luthi2007physical}. Furthermore, the use of different sound-wave polarizations permits to probe low-energy excitations in detail.

The usual interpretation of ultrasound data is based on a macroscopic picture of the elastic free energy for different strains. Within Landau's phenomenology, ultrasound has been widely used for the study of magnetic phase transitions such as in FeF$_2$ \cite{ikushima1971acoustic} and chromium \cite{muir1987magnetoelastic}. However, in order to investigate the effect of quantum fluctuations close to zero temperature, the use of a microscopic picture involving the interaction of phonons with low-energy magnetic excitations is inevitable. Recently, theoretical and experimental work has been devoted to build such approach for ultrasound data for frustrated magnets at low temperature, based on microscopic Heisenberg models and linear spin-wave theory \cite{kreisel2011elastic,bhattacharjee2011interplay}. Using microscopic models allows to take into account the crystal structure in a precise way, and in particular permits to compare different phonon polarizations without any additional free parameters. We used this approach to investigated magnetic-field dependence of magnetic and magnetoelastic properties of single-crystalline MnSc$_2$Se$_4$ samples at low temperatures. Providing a microscopic model based on the neutron-scattering results of Ref. \cite{guratinder2022magnetic}, we propose a magnetic excitation spectrum Ansatz for MnSc$_2$Se$_4$ at low temperature. We reproduce qualitatively the observed changes of the sound velocity and sound attenuation induced by the magnetic field and, in particular, give an interpretation of the observed difference between the measurements using longitudinal and transverse sound-wave polarizations.

\section{Experimental details}\label{sec2}

We grew single crystals of MnSc$_2$Se$_4$ using the chemical transport technique, as described in more detail in Ref. \cite{gao2017spiral}. We analyzed the composition with powder x-ray diffraction on a crushed single crystal, and by comparing the x-ray refinement and the measurements of magnetization versus temperature curve of the single crystals with a polycrystalline sample (Supplemental Material \cite{supp_mat}). For magnetization measurements, we selected a crystal of dimensions 1.0 mm $\times$ 1.0 mm $\times$ 2.0 mm. Since MnSc$_2$S$_4$ shows a very rich phase diagram with an extended skyrmion phase for magnetic fields applied along [111] \cite{gao2020fractional}, we performed our experiments using this field orientation. For ultrasound measurements, we selected a larger crystal of dimensions 1.3 mm $\times$ 3.5 mm $\times$ 2.0 mm, and polished on two faces perpendicular to the [111] axis.

We obtained high-field magnetization curves $M(H)$ up to $23$ T in pulsed magnetic fields at the Dresden High Magnetic Field Laboratory using the compensated pick-up coil system described in Refs. \cite{tsirlin2009exploring, skourski2011high}. We calibrated the absolute value of the magnetization by a low-field measurement using a commercial vibrating-sample magnetometer (VSM).

We performed ultrasound measurements utilizing the transmission pulse-echo technique with a phase-sensitive detection as described in Ref. \cite{luthi2007physical}. We attached LiNbO$_3$ transducers (36°-Y cut and 41°-X cut for longitudinal and for transverse mode, respectively) to the polished surfaces of the single crystal. We applied magnetic fields up to 17 T along the [111] axis in a superconducting magnet. We have set the ultrasound propagation direction $\textbf{k}$ parallel to the magnetic field $\textbf{k} \parallel \textbf{H} \parallel [111]$. We give the ultrasound frequencies and velocities at $4$ K for the two different acoustic modes in Table \ref{USv}.

\begin{table}[h!]
    \centering
     \caption{Ultrasound frequencies and velocities at $4$ K. $C_{11}$, $C_{12}$ and $C_{44}$ denote the elastic constants in Voigt notation.}
     \label{USv}
     \scalebox{1}{\begin{tabular}{ccccc} 
        \toprule%
        \midrule
         &  & Longitudinal & & Transverse \\
        &    & $\frac{1}{3}\left(C_{11} + 2C_{12} + 4C_{44}\right)$ & & $\frac{1}{3}\left(C_{11} - C_{12} + C_{44}\right)$ \\
         \midrule
         $f$ (MHz)& & 80  & &  65  \\
         $v$ (m/s)& & 3809 $\pm$ 200 & & 1809 $\pm$ 100 \\
         \midrule
         \bottomrule
     \end{tabular}}
\end{table}

\section{Results}\label{sec3}

In this section, we present the magnetization, sound-velocity and attenuation changes measured as a function of magnetic field in MnSc$_2$Se$_4$ single crystals.

\subsection{Magnetization}

\begin{figure}
\centering
\includegraphics[width=.93\linewidth]{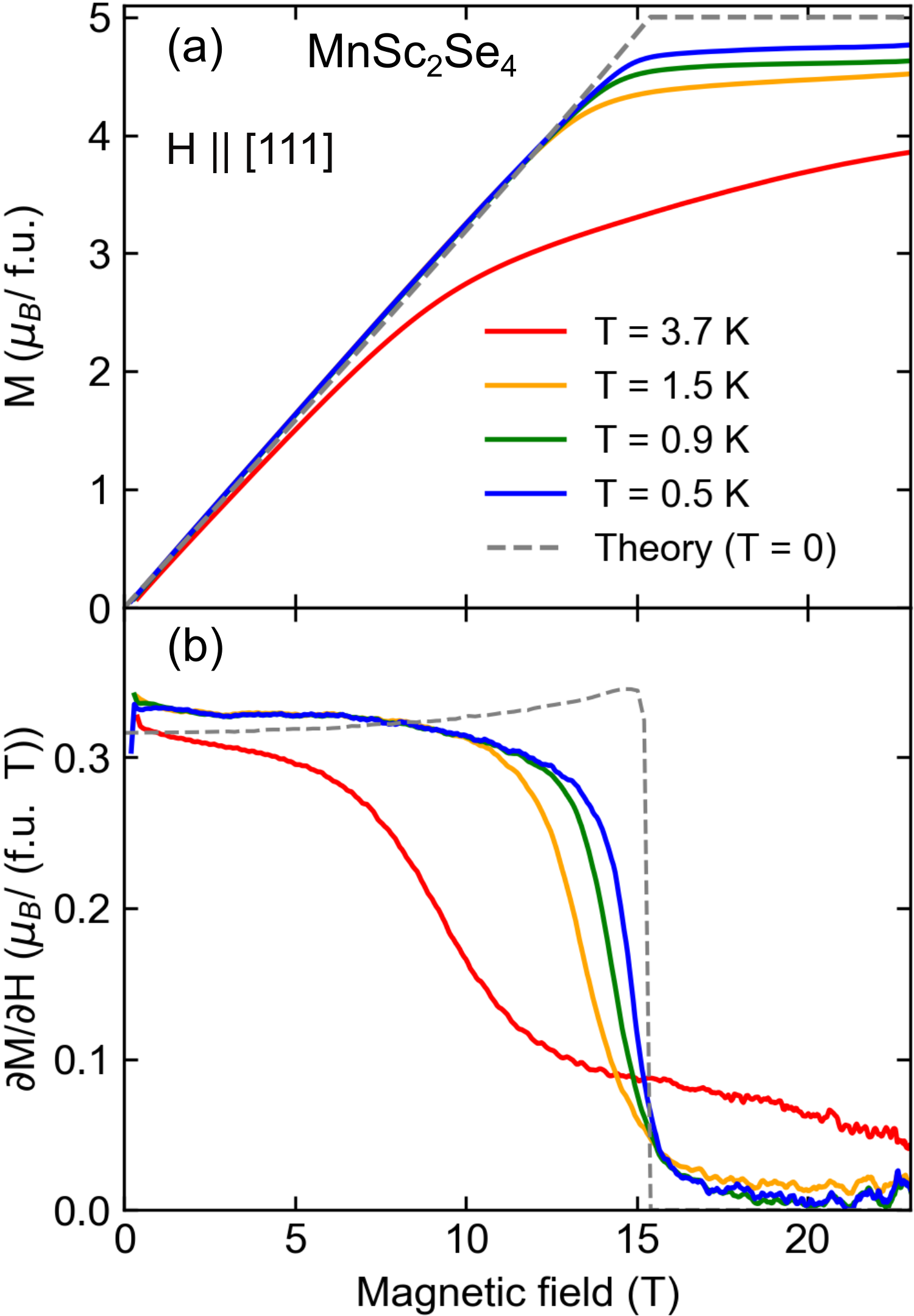}
\caption{Field-dependent (a) magnetization and (b) magnetization derivative of MnSc$_2$Se$_4$ for different temperatures, with the magnetic field applied along the [111] axis. Colored lines represent the experimental data while the grey dashed line corresponds to the calculation at $T = 0$ from the Heisenberg model presented in Section \ref{sec4}.}\label{fig1} 
\end{figure}

Figure \hyperref[fig1]{\ref*{fig1}} displays the magnetization of MnSc$_2$Se$_4$ and its derivative versus magnetic field applied along the [111] axis, measured at different temperatures. Below $1.5$ K, we observe a clear kink of the magnetization curve about 15.4 T, which indicates saturation as identified by the vanishing of $\partial M/ \partial H$ [Fig. \hyperref[fig1]{\ref*{fig1}(b)}]. This saturation suggests the presence of a polarized state. At $0.5$ K, the magnetic moment at saturation is about 4.75 $\mu_B$/f.u., which is smaller but close to the value $2S = 5$ $\mu_B$/f.u. expected for $S = 5/2$ Mn$^{2+}$ ions with quenched orbital moment \cite{kittel2005introduction}. A similar behavior has been observed in MnSc$_2$S$_4$ \cite{krimmel2006magnetic}. In Ref. \cite{krimmel2006magnetic}, the small magnetic moment in MnSc$_2$S$_4$ has been explained as originating from thermal fluctuations due to the low Néel temperature $T_N = 2.3$ K. This interpretation might be relevant for MnSc$_2$Se$_4$ too, since the Néel temperature of $2$ K is also small. At $T$ = 3.7 K, which is above the Néel temperature, the magnetization is reduced compared to the value at 1.5 K.

\subsection{Sound velocity and attenuation}

\begin{figure}
    \centering
    \includegraphics[width=.93\linewidth]{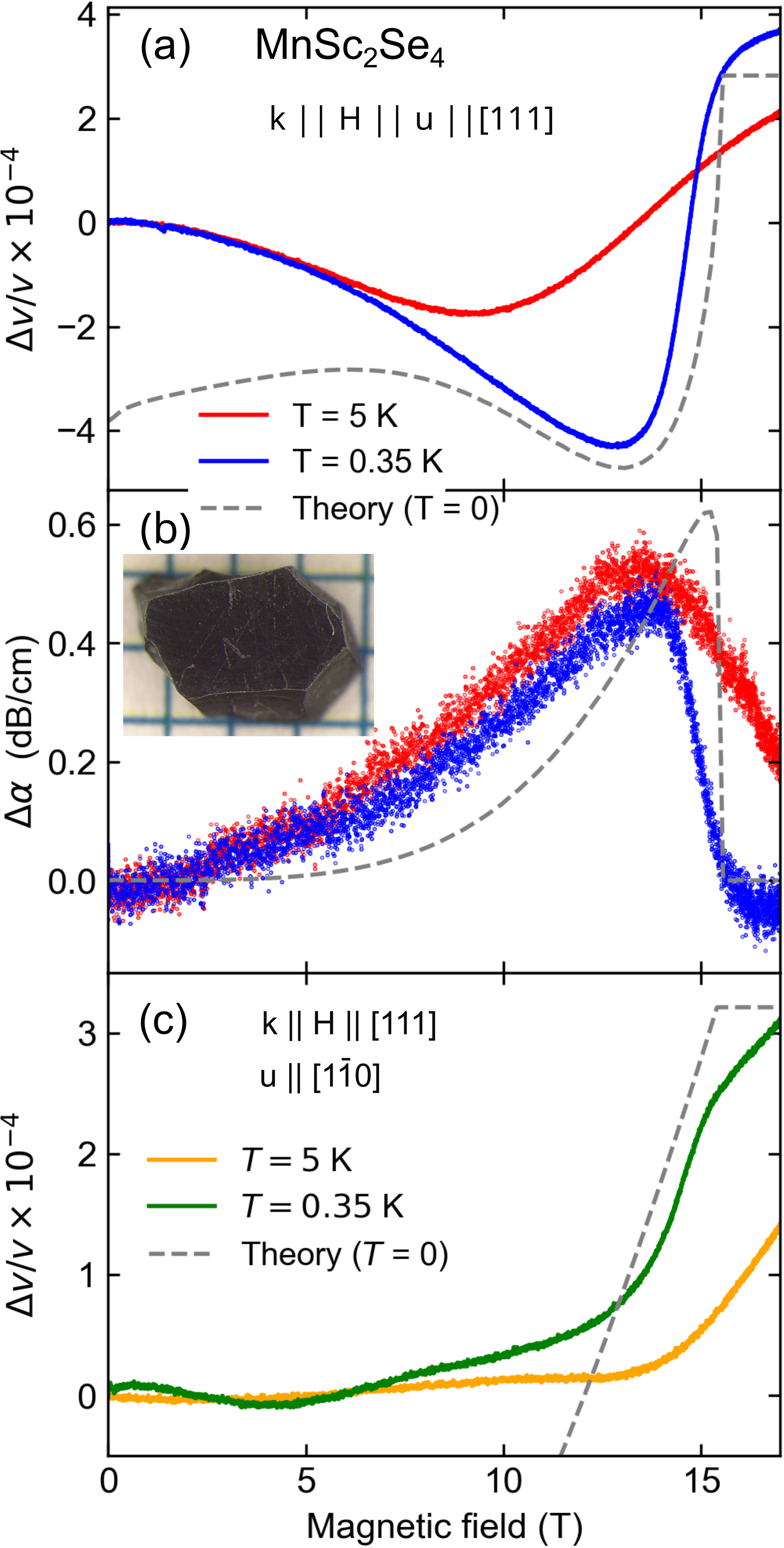}
    \caption{(a) Field-dependent sound-velocity change for longitudinal propagation and magnetic field applied along [111]. (b) Corresponding ultrasound-attenuation change. (c) Field-dependent sound-velocity change for transverse propagation with wave vector $\mathbf{k}$ and magnetic field along [111], and polarization $\mathbf{u}$ along $[1\bar{1}0]$. The inset in (b) shows the crystal used for
    the ultrasound measurements with polished [111] surface.
}\label{fig2} 
\end{figure}

Figure \hyperref[fig2]{\ref*{fig2}} shows the relative change of the sound velocity $\Delta v/v$ and attenuation $\Delta \alpha$ in MnSc$_2$Se$_4$ as a function of magnetic field, measured at different temperatures. No field-induced transition is observed, in accordance with the magnetization data. Instead a smooth softening of the longitudinal acoustic mode is detected [Fig. \hyperref[fig2]{\ref*{fig2}(a)}]. At 0.35 K, a minimum in the sound velocity occurs at $13$ T, which is below the saturation field. Close to the saturation field at about 15 T, the phonon softening is suppressed and we observe a hardening. Finally, for fields above $16$ T, the sound velocity seems to reach a constant value. At $5$ K, the minimum is shifted towards smaller fields, in agreement with the magnetization data where the kink is shifted towards smaller fields with increasing temperature. Furthermore, we detect a peak in the ultrasound attenuation just below the saturation field [Fig. \hyperref[fig2]{\ref*{fig2}(b)}]. This attenuation peak marks an increase of the energy dissipation in the system and has been reported for other antiferromagnetic materials close to the saturation field, such as CsCuCl$_4$ \cite{kreisel2011elastic}. The shift of the attenuation peak towards lower fields at the higher temperature of 5 K is less pronounced than the corresponding shift of the sound-velocity minimum, because the attenuation peak should occur close to the inflection point of the sound-velocity variation. We interpret this as a result of the reduction of the magnetic correlations due to thermal fluctuations, in accordance with the magnetization measurements.

Remarkably, we do not observe any significant phonon softening for the transverse acoustic mode [Fig. \hyperref[fig2]{\ref*{fig2}(c)}]. For this mode, there occurs only a hardening close the saturation field. Furthermore, we remark that this hardening is about $3 \times 10^{-4}$ at 17 T, comparable to the hardening observed for the longitudinal mode. A small anomaly is observed at T = 0.35 K and magnetic field lower than 5 T. Its origin remains unclear, as it is not observed in the longitudinal mode, nor in the magnetization data.

For both longitudinal and transverse acoustic modes, the sound-velocity variation is less pronounced at 5 K, compared to the variation at 0.35 K. This is in accordance with the reduced magnetization observed above the Néel temperature (Fig. \hyperref[fig1]{\ref*{fig1}}) and suggests that at low temperature the major contribution to the elastic property changes of MnSc$_2$Se$_4$ originates in the spin-lattice coupling. Thus, the dominating energy scales for the elastic properties correspond to those of the magnetic subsystem, namely the Néel temperature $T_N = 2$ K and the saturation field $\mu_0 H_{sat} = 15.4$ T. Furthermore, above the saturation field, the [111] axis becomes the hard magnetic axis and induces a hardening of the elastic modes for propagation along [111]. A similar effect has been reported in an experimental study of the cone state of MnSi and explained theoretically using free-energy calculations \cite{plumer1982magnetoelastic,nii2014elastic}.

\section{Theory}\label{sec4}

In this section, we present a microscopic model which reproduces qualitatively the observed sound-velocity and attenuation changes in MnSc$_2$Se$_4$ upon varying the magnetic field. This model follows closely the methods of Ref. \cite{kreisel2011elastic} for the description of sound-velocity changes in the cone state of CsCuCl$_4$, and of Ref. \cite{bhattacharjee2011interplay} for the frustrated magnet CdCr$_2$O$_4$. We define a microscopic Hamiltonian, reproducing the dynamics of spin degrees of freedom $\mathcal{H}_{s}$ and lattice degrees of freedom $\mathcal{H}_{l}$, together with a spin-lattice interaction term $\mathcal{H}_{sl}$, which can be written as:
\begin{equation}\label{eq1}
    \begin{split}
            \mathcal{H} &= \mathcal{H}_{s} + \mathcal{H}_{l} + \mathcal{H}_{sl}.
    \end{split}{}
\end{equation}{}
Each term will be precised hereafter. In the first part, we propose an Ansatz of a cone state for the magnetic order in MnSc$_2$Se$_4$ in applied magnetic field. We adjust the microscopic model parameter in order to reproduce the saturation field observed experimentally and determine the spin-wave spectrum using linear spin-wave theory. In the second part, we derive the spin-lattice coupling based on the exchange-striction mechanism. This permits us to determine the phonon self-energy and the associated variations of the sound velocity and attenuation in the third part. Finally, we compare the theoretical results to the ultrasound experiments in the fourth part.

\subsection{Spin dynamics $\mathcal{H}_{s}$}\label{Hs}

The dynamics in MnSc$_2$Se$_4$ at low temperature is believed to be dominated by fluctuations of magnetic moments from Mn$^{2+}$ ions, which will be represented by a Heisenberg Hamiltonian. The Mn$^{2+}$ ions occupy a diamond lattice, represented in Fig. \hyperref[fig3]{\ref*{fig3}}. The neutron-scattering data on MnSc$_2$Se$_4$ \cite{guratinder2022magnetic} suggest to consider magnetic couplings up to the third neighbor, with the values $J_1/k_B = -0.24$ K, $J_2/k_B = 0.37$ K, and $J_3/k_B = 0.072$ K. Denoting $a$ and $b$ the two face-centered cubic (fcc) sublattices of the diamond lattice of the Mn$^{2+}$ ions and $\mathbf{S}_i^a$, $\mathbf{S}_i^b$ the associated magnetic moments, the spin Hamiltonian $\mathcal{H}_s$ is written as: 
\begin{figure}
    \centering
    \includegraphics[width=\linewidth]{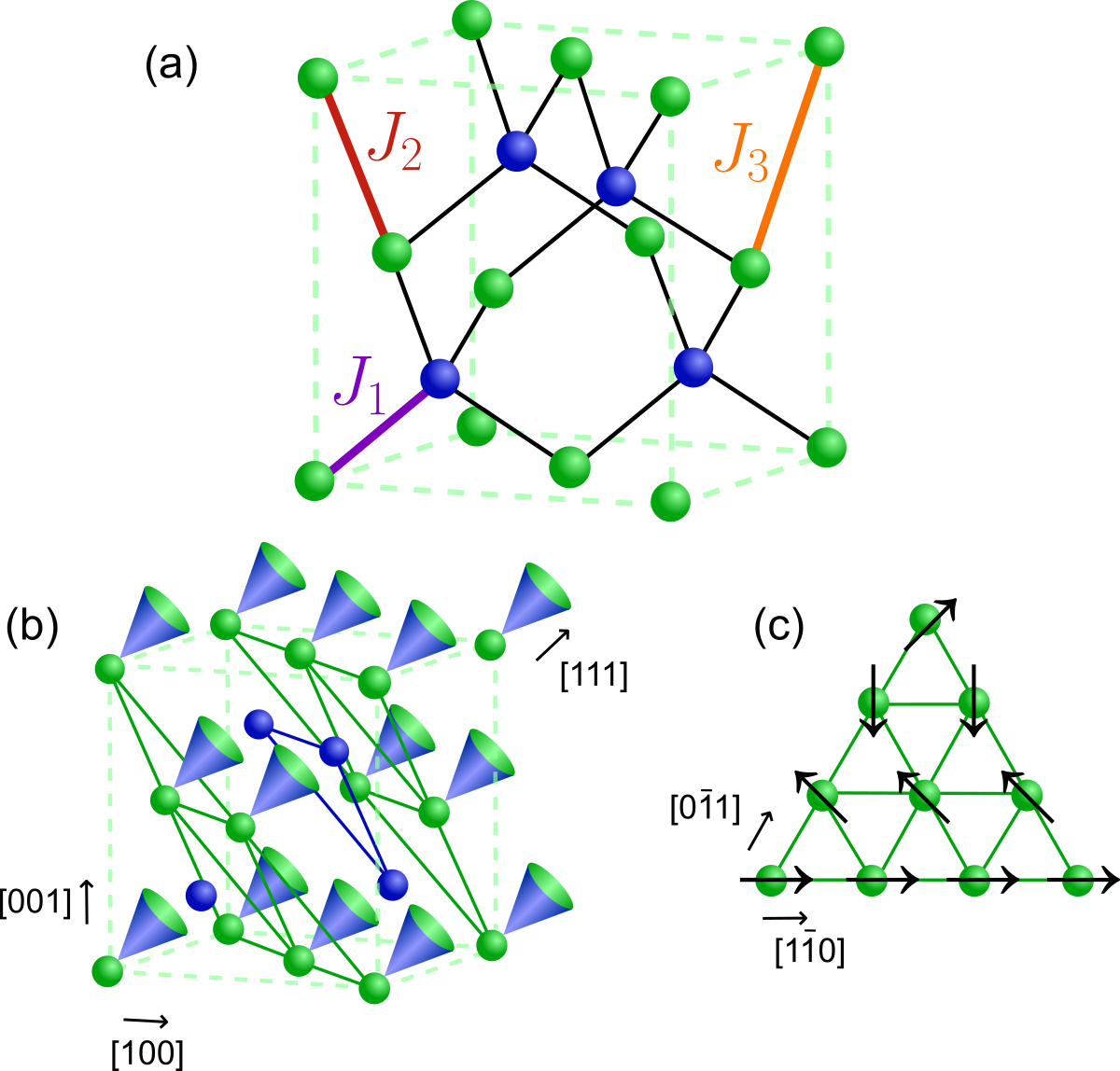}
    \caption{(a) The diamond lattice of the Mn$^{2+}$ ions in MnSc$_2$Se$_4$, consisting of two fcc sublattices (blue and green), together with the couplings $J_1$, $J_2$, and $J_3$. (b) Cone state for MnSc$_2$Se$_4$, with the opening of the cone along the [111] axis. The crystal structure is represented as a superposition of triangular lattices perpendicular to [111], which shows the geometric frustration induced by the coupling $J_2$. (c) Magnetic order on one triangular lattice perpendicular to [111], which corresponds to the magnetic ordering vector $\mathbf{q}_0 = [\frac{3\pi}{2a_0},-\frac{3\pi}{2a_0},0]$ observed in neutron diffraction \cite{guratinder2022magnetic}. Between neighboring [111] planes, there is an additional rotation of the magnetic structure by $\frac{3\pi}{4}$.}\label{fig3}
\end{figure}
\begin{equation}\label{eq2}
    \begin{split}
      \mathcal{H}_{s} = &\sum_{i\delta_1}J_{1}\mathbf{S}_i^a \cdot \mathbf{S}_{i+\delta_1}^b +\frac{1}{2}\sum_{i\delta_2}J_{2}\left(\mathbf{S}_i^a \cdot \mathbf{S}_{i+\delta_2}^a + \mathbf{S}_i^b \cdot \mathbf{S}_{i+\delta_2}^b\right) \\
       + &\sum_{i\delta_3}J_{3}\mathbf{S}_i^a \cdot \mathbf{S}_{i+\delta_3}^b  - \sum_i \mathbf{h}\cdot \left(\mathbf{S}_i^a+\mathbf{S}_i^b\right),
    \end{split}{}
\end{equation}{}
where $i$ runs over the sites of one fcc sublattice, $\delta_1$, $\delta_2$, and $\delta_3$ join the site $i$ with its first, second, and third nearest neighbors, respectively, as represented in Fig. \hyperref[fig3]{\ref*{fig3}}. $\mathbf{h} = g\mu_B \mathbf{H}$ is related to the external magnetic field $\mathbf{H}$, where we assumed the value $g=2$ for the Landé factor of the Mn$^{2+}$ ion and with $\mu_B$ the Bohr magneton. It has been proposed in Ref. \cite{lee2008theory} that in the presence of external magnetic field, the ground state of this Heisenberg model is a conical helix, with the helical ordering vector perpendicular to the magnetic field direction. For a magnetic field along [111], we will consider a state of ordering vector $\mathbf{q}_0 = [q_0, -q_0, 0]$. In order to determine the spin-wave spectrum associated to this conical helix, we use a locally-rotating reference frame, as done in Ref. \cite{kreisel2011elastic}. We first define three orthonormal basis vectors: $\mathbf{l}_0 \parallel$ [111] in the direction of the magnetic field, and $\mathbf{l}_1 \parallel$ $[1\bar{1}0]$, $\mathbf{l}_2 \parallel$ $[11\bar{2}]$ perpendicular to it. Then, we define the reference frame by:
\begin{equation}\label{eq3}
    \begin{split}
        \mathbf{m}_i^{\alpha} &= \cos\theta\left[\cos(\mathbf{q}_0\cdot\mathbf{r}_i^\alpha)\mathbf{l}_1 + \sin(\mathbf{q}_0\cdot\mathbf{r}_i^\alpha)\mathbf{l}_2\right] + \sin\theta\mathbf{l}_0, \\
        \mathbf{e}_i^{\alpha 1} &=\sin(\mathbf{q}_0\cdot\mathbf{r}_i^\alpha)\mathbf{l}_1 - \cos(\mathbf{q}_0\cdot\mathbf{r}_i^\alpha)\mathbf{l}_2, \\
        \mathbf{e}_i^{\alpha2} &= \sin\theta\left[\cos(\mathbf{q}_0\cdot\mathbf{r}_i^\alpha)\mathbf{l}_1 + \sin(\mathbf{q}_0\cdot\mathbf{r}_i^\alpha)\mathbf{l}_2\right]  - \cos\theta\mathbf{l}_0,
    \end{split}
\end{equation}
where $\mathbf{m}_i^{\alpha}$ points in the direction of the local magnetization at the position $\mathbf{r}_i^\alpha$, which corresponds to the site $i$ of the sublattice $\alpha = a,b$. $\mathbf{e}_i^{\alpha 1}$, $ \mathbf{e}_i^{\alpha 2}$ are two mutually perpendicular transverse vectors that permit to construct an orthonormal basis, and $\theta$ is the opening angle of the conical helix. Defining $\mathbf{e}_i^{\alpha\pm} = \mathbf{e}_i^{\alpha1} \pm i \mathbf{e}_i^{\alpha2}$, the magnetic moment of a given Mn$^{2+}$ ion can be written in the locally rotating reference frame as:
\begin{equation}\label{eq4}
    \begin{split}
        \mathbf{S}_i^{\alpha} &=  S_i^{\alpha\parallel}\mathbf{m}_i^\alpha + \frac{1}{2}S_i^{\alpha+}\mathbf{e}_i^{\alpha-} + \frac{1}{2}S_i^{\alpha-}\mathbf{e}_i^{\alpha+}.
    \end{split}
\end{equation}
The spin operators are then written with the Holstein-Primakoff transformation \cite{holstein1940field}: $S_{i}^{a\parallel} = S - a_i^\dagger a_i$, $S_{i}^{a+} = \sqrt{2S}a_i$, $S_{i}^{a-} = \sqrt{2S}a_i^\dagger$ and $S_{i}^{b\parallel} = S - b_i^\dagger b_i$, $S_{i}^{b+} = \sqrt{2S}b_i$, $S_{i}^{b-} = \sqrt{2S}b_i^\dagger$. With this formulation, we obtain linear and quadratic terms in magnon operators. The ground-state energy is minimized for:
\begin{equation}\label{eq5}
    \begin{split}
        h &= S\sin\theta \sum_{\delta_\alpha} J_{\delta_\alpha} \left[1-\cos(\mathbf{q}_0 \cdot \boldsymbol{\delta}_\alpha)\right],
    \end{split}
\end{equation}
which is equivalent to a vanishing linear term in magnon operators. Thus, the value of the magnetic field is directly related to the opening of the helix, and in the following we will use $\sin \theta = H/H_c$, where $H_c$ is the saturation field. The remaining Heisenberg Hamiltonian of Eq. \hyperref[eq2]{(\ref*{eq2})} is now quadratic in magnon operators, and using the Fourier transform with an equal number of $a$ and $b$ sites $N$: $a_i = 1/\sqrt{N}\sum_q e^{i\mathbf{q} \cdot \mathbf{r}_i} a_q$, $b_i= 1/\sqrt{N}\sum_q e^{i\mathbf{q} \cdot \mathbf{r}_i} b_q$, we obtain:

\

\begin{figure}[h]
    \centering
    \includegraphics[width=.96\linewidth]{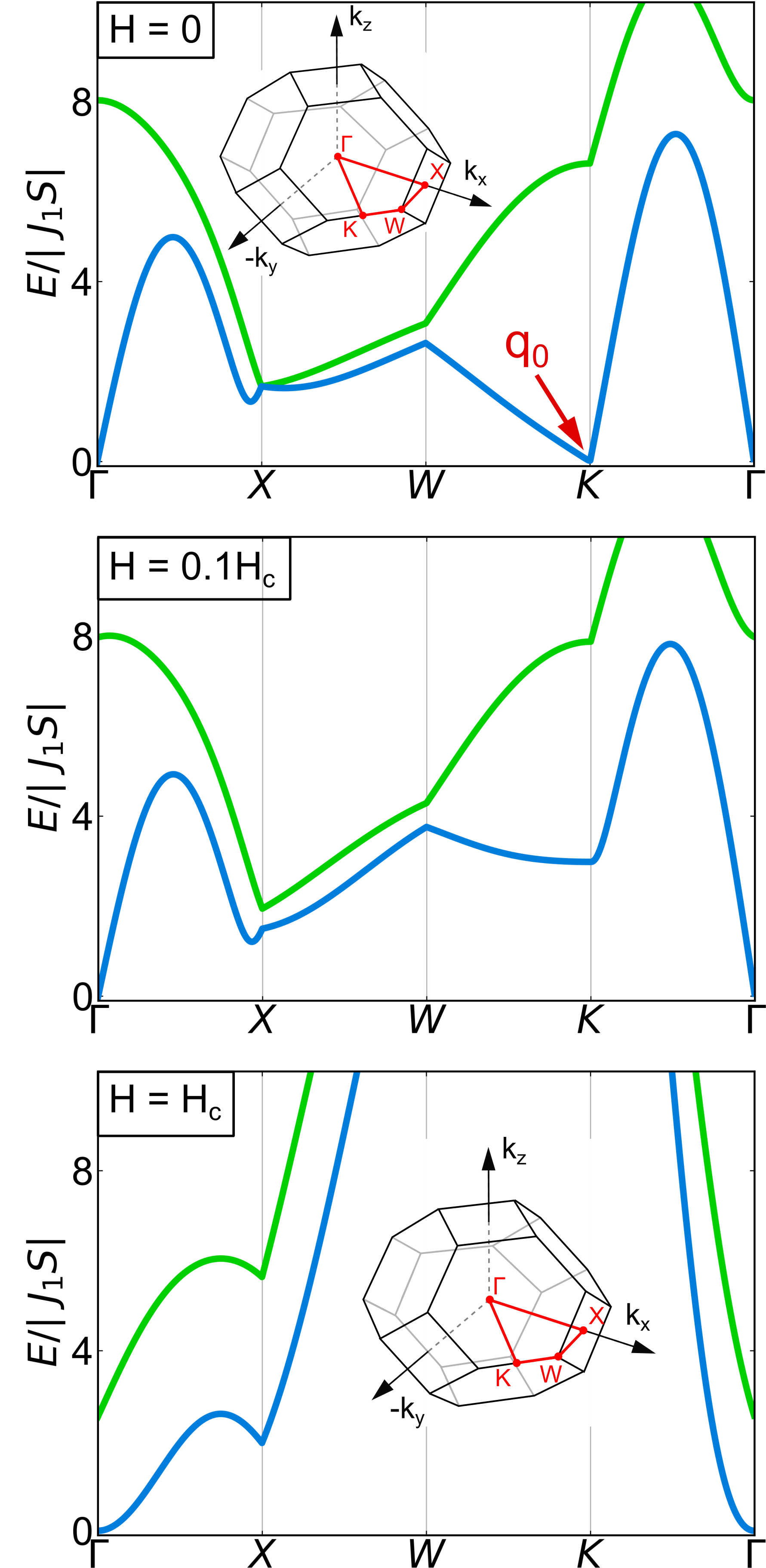}
    \caption{Magnon dispersion for the Heisenberg Hamiltonian of Eq. \hyperref[eq6]{(\ref*{eq6})} for different values of the external magnetic field: $H=0$ (a), $H=0.1H_c$ (b), and $H=H_c$ (c). The high-symmetry contour of the diamond-lattice Brillouin zone is defined by $\Gamma = (0,0,0)$, $X = (\frac{2\pi}{a_0},0,0)$, $W= (\frac{2\pi}{a_0},-\frac{\pi}{a_0},0)$ and $K= (\frac{3\pi}{2a_0},-\frac{3\pi}{2a_0},0)$, where $a_0$ is the lattice spacing. The values of the couplings have been set to $J_2/|J_1| = 1.54$, $J_3/|J_1| = 0.3$.}\label{fig4}
\end{figure}

\begin{widetext}
    \begin{equation}\label{eq6}
        \begin{split}
            \mathcal{H}_{s} &= E_0 + \sum_{q}A_q \left(a_q^\dagger a_q + b_q^\dagger b_q \right)+ \sum_{q}\frac{1}{2}B_q \left(a_q^\dagger a_{-q}^\dagger + a_q^\dagger a_{-q} + b_q^\dagger b_{-q}^\dagger + b_q b_{-q} \right), \\
            & \ \ \ \ \ \ \ \ + \sum_{q} \left(C_qa_q^\dagger b_q + C_q^* a_q b_q^\dagger  \right) + \sum_{q} \left(D_q a_q^\dagger b_{-q}^\dagger + D_q^*a_q b_{-q} \right),\\
            E_0 &= NS^2 \sum_{\delta_\alpha = \delta_1,\delta_2, \delta_3}J_{\delta_\alpha}\left[ (1+\sin\theta^2) \cos(\mathbf{q}_0 \cdot \boldsymbol{\delta}_\alpha)-\sin\theta^2\right],\\
            A_q &= -S \sum_{\delta_\alpha= \delta_1,\delta_2, \delta_3} J_{\delta_\alpha}\cos(\mathbf{q}_0 \cdot \boldsymbol{\delta}_\alpha) + \frac{SJ_2}{2} \sum_{\delta_2} \left[ \left(1 + \sin\theta^2 \right)\cos(\mathbf{q}_0 \cdot \boldsymbol{\delta}_2) + \cos\theta^2 +2i\sin\theta\sin(\mathbf{q}_0 \cdot \boldsymbol{\delta}_{2}) \right]e^{i\mathbf{q} \cdot \boldsymbol{\delta}_{2}},\\
            B_q &= \frac{SJ_2}{2} \sum_{\delta_2} \cos\theta^2\left[\cos(\mathbf{q}_0 \cdot \boldsymbol{\delta}_2) - 1  \right]\cos(\mathbf{q} \cdot \boldsymbol{\delta}_2),\\
            C_q &= \frac{S}{2} \sum_{\delta_{\alpha'} = \delta_1, \delta_3} J_{\delta_{\alpha'}}\left[\left(1 + \sin\theta^2 \right)\cos(\mathbf{q}_0 \cdot \boldsymbol{\delta}_{\alpha'})+\cos\theta^2 +2i\sin\theta\sin(\mathbf{q}_0 \cdot \boldsymbol{\delta}_{\alpha'}) \right]e^{i\mathbf{q} \cdot \boldsymbol{\delta}_{\alpha'}},\\
            D_q &= \frac{S}{2} \sum_{\delta_{\alpha'} = \delta_1, \delta_3} J_{\delta_{\alpha'}}  \cos\theta^2\left[\cos(\mathbf{q}_0 \cdot \boldsymbol{\delta}_{\alpha'}) - 1  \right]e^{i\mathbf{q} \cdot \boldsymbol{\delta}_{\alpha'}},
        \end{split}{}
    \end{equation}{}
\end{widetext}
with $\mathbf{q}_0 = [\frac{3\pi}{2a_0},-\frac{3\pi}{2a_0},0]$, and $a_0$ is the lattice spacing.
Because the coupling $J_1$ connects the center of a manganese tetrahedron to its vertices [see Fig. \hyperref[fig3]{\ref*{fig3}(a)}], there is no inversion symmetry on a given manganese site and we did not obtain an analytic expression for the Bogoliubov transformation in the general case. We diagonalized the Hamiltonian of Eq. \hyperref[eq6]{(\ref*{eq6})} numerically, using the algorithm of Colpa \cite{colpa1978diagonalization} described in Supplemental Material \cite{supp_mat}. We obtain:
\begin{equation}\label{eq7}
    \begin{split}
      \mathcal{H}_{s} &= E_0 +\Delta E + \sum_{q} \left(E^+_q \alpha_q^\dagger \alpha_q + E^-_q \beta_q^\dagger \beta_q \right),\\
      \Delta E & = - \sum_q A_q + \frac{1}{2} \sum_q \left( E_q^+ + E_q^-\right),
    \end{split}{}
\end{equation}{}
with the results shown in Fig. \hyperref[fig4]{\ref*{fig4}}. At zero magnetic field, two soft modes are present at $\mathbf{q} = [0,0,0]$ and $\mathbf{q} = \mathbf{q}_0$, which correspond to the $\Gamma$ and $K$ points on the high-symmetry contour, respectively [Fig. \hyperref[fig4]{\ref*{fig4}(a)}]. When applying a magnetic field, a planar anisotropy is induced \cite{mourigal2010field} and a gap opens at the $K$ point [Fig. \hyperref[fig4]{\ref*{fig4}(b)}]. Finally, for $H\rightarrow H_c$, the magnon velocity $v_{mag}$ becomes zero, leading to a parabolic dispersion around the $\Gamma$ point, typical for polarized systems [Fig. \hyperref[fig4]{\ref*{fig4}(c)}]. In the $q\rightarrow 0$ limit, the Bogoliubov transformation can be done analytically (Supplemental Material \cite{supp_mat} and reference \cite{rastelli1990frustration} therein), and we show that the magnon velocity is a decreasing function of the magnetic field, being maximum at $H = 0$.

The knowledge of the spin-wave spectrum allows us to calculate the magnetization and its derivative, using $M = - \frac{1}{2N}\frac{\partial (E_0 + \Delta E)}{\partial H}$ (Supplemental Material \cite{supp_mat} and reference \cite{ zhitomirsky1998magnetization} therein). We can first determine the saturation field with Eq. \hyperref[eq5]{(\ref*{eq5})}. Using $S = 5/2$, $g=2$, and the values of Ref. \cite{guratinder2022magnetic} for the $J$ couplings, we obtain a saturation field of $11.1$ T, which is somewhat lower than the experimental value. Some differences between predicted values of the $J$ couplings are already discussed in the case of MnSc$_2$S$_4$ \cite{gao2020fractional,iqbal2018stability}, and might be relevant here. Keeping the ratios $J_{\delta_\alpha}/|J_1|$ fixed in order to reproduce $\mathbf{q}_0 = [\frac{3\pi}{2a_0},-\frac{3\pi}{2a_0},0]$ observed experimentally, our data suggest a bigger energy scale $|J_1|/k_B = 0.33$ K, leading to $\mu_0 H_{sat} = 15.4$ T. Furthermore, this new value produced a theoretical Curie-Weiss temperature of $\frac{S(S+1)}{3 k_B} \sum_{\delta_\alpha}J_{\delta_\alpha} = -17.4$ K \cite{yosida1996theory}, which is close to the experimental value of $\Theta_{CW} = -18.4$ K.
Using these refined values of the $J$ couplings, we calculate $M$ and $\partial M/ \partial H$ (grey dashed lines in Fig. \hyperref[fig1]{\ref*{fig1}(b)}). The quantum correction to the energy $\Delta E$ generates an upturn of $\partial M/\partial H$ with increasing magnetic field, which should be small for large values of $S$ and is not observed experimentally. We also calculate the magnon velocity at zero field $v_{mag} = 470$ m/s. This velocity is, thus, roughly one order of magnitude smaller than the phonon velocity $v$ measured experimentally (Table \ref{USv}), which is a consequence of the small magnitude of the magnetic interactions in MnSc$_2$Se$_4$ with $T_N = 2$ K and $\Theta_{CW} = 18.4$ K.
\subsection{Lattice dynamics $\mathcal{H}_{l}$ and exchange-striction coupling $\mathcal{H}_{sl}$}

In Heisenberg models, the spin-lattice coupling can be introduced by evaluating the variation of the exchange parameters with respect to atomic displacements \cite{stern1965thermal, tachiki1974effect}. Atomic displacements are taken into account by writing the position of a given Mn$^{2+}$ ion as $\mathbf{R}_i^\alpha = (\mathbf{R}_i^\alpha)^0 + \mathbf{Q}_i^\alpha$, with the equilibrium position $(\mathbf{R}_i^\alpha)^0$ and the displacement $\mathbf{Q}_i^\alpha$. Writing the bond $\boldsymbol{\delta}_\alpha = \mathbf{R}_i^\alpha - \mathbf{R}_{i+\delta}^{\alpha'}$, the Taylor expansion of the coupling parameters $J_{\delta_{\alpha}}$ around the equilibrium bond distance gives:
\begin{equation}\label{eq8}
    \begin{split}
        J_{\delta_\alpha} \approx J_{\delta_\alpha}^0 &+ \sum_{\mu} \frac{\partial J_{\delta_\alpha}}{\partial R^\mu}(Q_i^{\alpha\mu}-Q_{i+\delta}^{\alpha'\mu}) \\
    &+ \frac{1}{2} \sum_{\mu,\nu} \frac{\partial^2 J_{\delta_\alpha}}{\partial R^\mu\partial R^\nu}(Q_i^{\alpha\mu}-Q_{i+\delta}^{\alpha'\mu})(Q_i^{\alpha\nu}-Q_{i+\delta}^{\alpha'\nu}),
    \end{split}{}
\end{equation}{}
where $\mu,\nu = x,y,z$ are Cartesian coordinates. The atomic displacement $\mathbf{Q}_i^\alpha$ at atomic position $i$ is then quantized with the usual phonon destruction and creation operators $c_{k\lambda}$ and $c_{k\lambda}^\dagger$, of momentum $k$ and polarization $\lambda$ \cite{mahan2000many} :
\begin{equation}\label{eq9}
    \begin{split}
        \mathbf{Q}_i^\alpha &= i \sum_{k\lambda} \sqrt{\frac{1}{2M_0N\omega_{k\lambda}^0}} \boldsymbol{\epsilon}_{k\lambda} (c_{k\lambda}+c_{-k\lambda}^\dagger)e^{i\mathbf{k}\cdot(\mathbf{R}_i^\alpha)^0},
    \end{split}{}
\end{equation}{}
where $M_0$ denotes the mass of the Mn$^{2+}$ ion, $N$ the number of sites in one fcc sublattice,  $\boldsymbol{\epsilon}_{k\lambda} = -\boldsymbol{\epsilon}_{-k\lambda}$ the polarization vector, and $\omega_{k\lambda}^0$ the energy of the phonon excitation. In the following, we will consider each phonon polarization independently, and we thus drop the index $\lambda$. Taking the dynamics of the phonons to be harmonic and using the expansion of Eq. \hyperref[eq8]{(\ref*{eq8})} in the spin Hamiltonian $\mathcal{H}_s$ of Eq. \hyperref[eq2]{(\ref*{eq2})}, the lattice Hamiltonian and the spin-lattice coupling can be written as:
\begin{equation}\label{eq10}
    \begin{split}
       \mathcal{H}_{l} &= \sum_{k} \omega_{k}^0 c_{k}^\dagger c_{k},\\
            \mathcal{H}_{sl} &= \sum_k U_k^1 C_k + \frac{1}{2}\sum_{kk'} U_{kk'}^2 C_kC_{k'},
    \end{split}{}
\end{equation}{}
where $C_k = c_k + c_{-k}^\dagger$ is the phonon displacement operator. $U^1_k$ and $U^2_{kk'}$ are given by:
\begin{equation}\label{eq11}
    \begin{split}
        U_k^1 &= \frac{1}{\sqrt{2M_0N\omega_{k}^0}} \sum_{i\delta_{\alpha}} \mathbf{S}_i^\alpha\cdot\mathbf{S}_{i+\delta_{\alpha}}^{\alpha'} \left(e^{i\mathbf{k}\cdot\mathbf{R}_i^\alpha}-e^{i\mathbf{k}\cdot\mathbf{R}_{i+\delta_{\alpha}}^{\alpha'}}\right)\\
        & \ \ \ \ \times \left(i \sum_\mu \epsilon_k^\mu \frac{\partial J_{\delta_\alpha}}{\partial R^\mu}\right), \\ 
        U_{kk'}^2 &=  \frac{1}{2M_0N\sqrt{\omega_{k}^0\omega_{k'}^0}}\sum_{i\delta_{\alpha}} \mathbf{S}_i^\alpha\cdot\mathbf{S}_{i+\delta_{\alpha}}^{\alpha'} \left(e^{i\mathbf{k}\cdot\mathbf{R}_i^\alpha}-e^{i\mathbf{k}\cdot\mathbf{R}_{i+\delta_{\alpha}}^{\alpha'}}\right)\\
        & \ \ \ \ \times\left(e^{i\mathbf{k}'\cdot\mathbf{R}_i^\alpha}-e^{i\mathbf{k'}\cdot\mathbf{R}_{i+\delta_{\alpha}}^{\alpha'}}\right) \left(i^2 \sum_{\mu\nu} \epsilon_k^\mu\epsilon_{k'}^\nu \frac{\partial^2 J_{\delta_\alpha}}{\partial R^\mu\partial R^\nu}\right).
    \end{split}{}
\end{equation}{}

Finally, upon replacing the spin operators $\mathbf{S}_i$ by their Holstein-Primakoff expansions, we obtain a system of interacting phonons and magnons, which represent the interaction of the lattice degrees of freedom and the spin-wave fluctuations.

The key ingredient of the spin-lattice coupling is the derivative of the exchange parameters $\partial J_{\delta_\alpha}/\partial R^\mu$. A simple ansatz is an exponential dependence of the exchange interaction $J_{\delta_\alpha}  = J_{\delta_\alpha}^0 \text{exp}(-\sqrt{M}_0 \xi |R_{i}^\alpha-R_{i+\delta}^{\alpha'}|)$, as used to explain the elastic properties of the frustrated magnet CdCr$_2$O$_4$ \cite{bhattacharjee2011interplay}, for example. In that case, the derivatives are easily evaluated: 
\begin{equation}\label{eq12}
    \begin{split}
       \frac{1}{\sqrt{M_0}} \left(i \sum_\mu \epsilon_k^\mu \frac{\partial J_{\delta_\alpha} }{\partial R^\mu}\right) &= -i\xi J^0_{\delta_\alpha} (\boldsymbol{\delta}_\alpha \cdot \boldsymbol{\epsilon}_k), \\
        \frac{1}{M_0}\left(i^2 \sum_{\mu\nu} \epsilon_k^\mu\epsilon_{k'}^\nu \frac{\partial^2 J_{\delta_\alpha} }{\partial R^\mu\partial R^\nu}\right) &= -\xi^2 J^0_{\delta_\alpha} (\boldsymbol{\delta}_\alpha \cdot \boldsymbol{\epsilon}_k) (\boldsymbol{\delta}_\alpha \cdot \boldsymbol{\epsilon}_{k'}),
    \end{split}{}
\end{equation}{}
and the spin-lattice coupling is then characterized by a single parameter $\xi$. In the following, we replace the notation $J_{\delta_\alpha}^0$ by $J_{\delta_\alpha} = J_1, J_2, J_3$.

\subsection{Phonon self-energy}

\begin{figure}
    \centering
    \includegraphics[width=\linewidth]{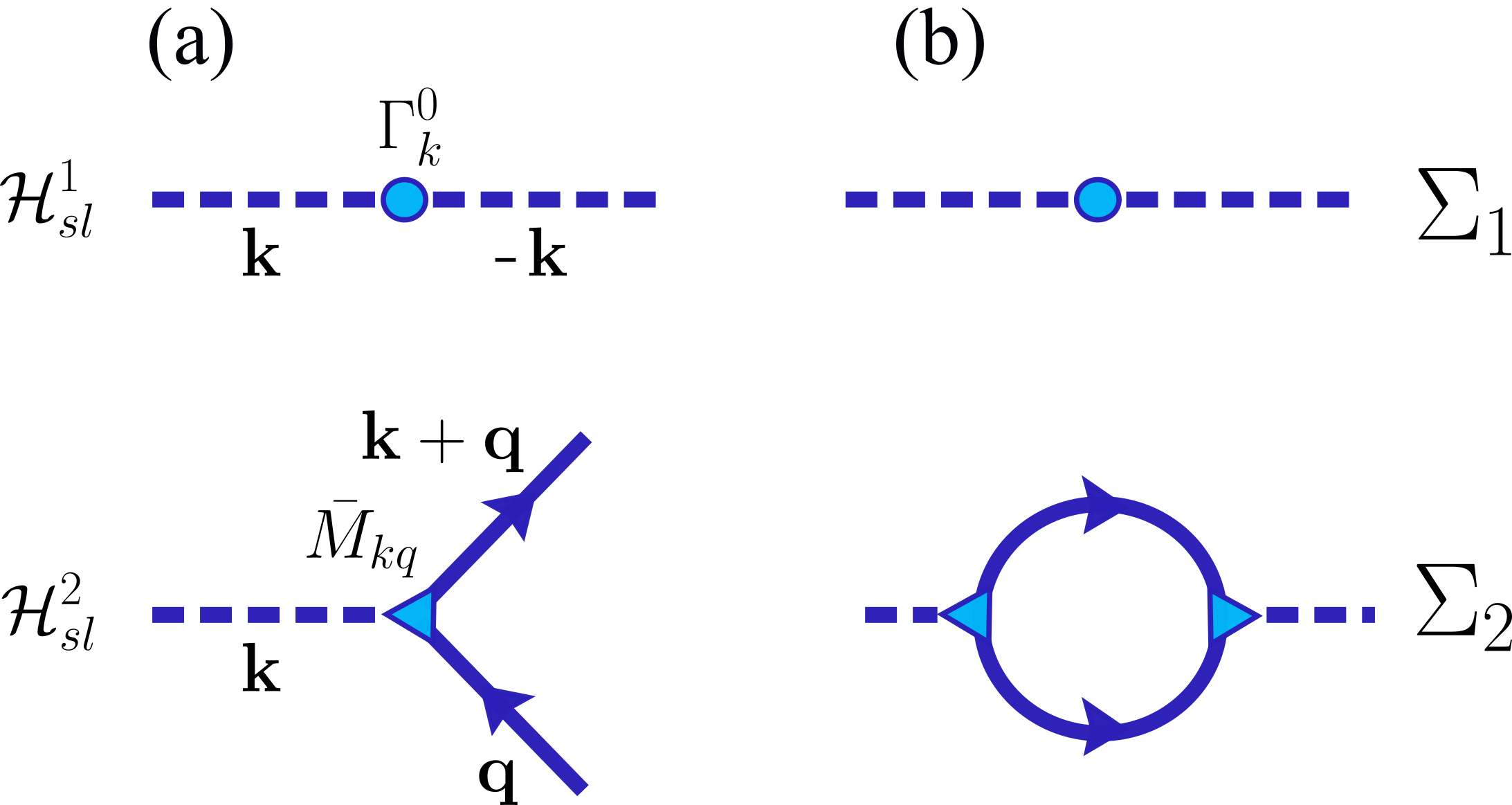}
    \caption{(a) Interaction terms induced by the spin-lattice coupling, as written in Eq. \hyperref[eq13]{(\ref*{eq13})}. Phonon propagators are represented by dashed lines, and magnon propagators by full lines. (b) Associated contribution to the phonon self-energy at zero temperature.}\label{fig5}
\end{figure}

\begin{figure}
    \centering
    \includegraphics[width=\linewidth]{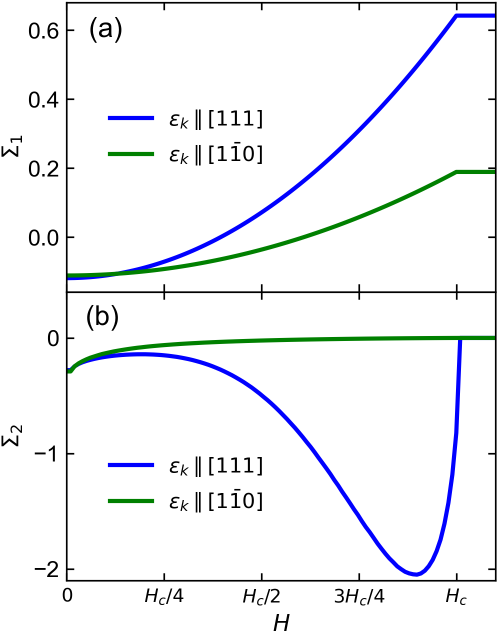}
    \caption{(a) Phonon self-energy contribution from the constant background $\Sigma_1$, in the case of longitudinal and transverse phonon polarization. (b) Phonon self-energy contribution from the magnon-pair creation bubble $\Sigma_2$, in the case of longitudinal and transverse phonon polarizations.}\label{fig6}
\end{figure}

In order to evaluate the phonon self-energy, we use the Bogoliubov-rotated basis for the magnon operators, denoting $\Psi_q^\dagger = (\beta_q^\dagger, \alpha_q^\dagger , \alpha_{-q},  \beta_{-q})$. In this basis, the magnon energy is written as $\left(\bar{E}_{q}\right)_{i,j} = E_q^-\delta_{i,1}\delta_{j,1} + E_q^+\delta_{i,2}\delta_{j,2} + E_{-q}^+\delta_{i,3}\delta_{j,3} + E_{-q}^-\delta_{i,4}\delta_{j,4}$. In Supplemental Material \cite{supp_mat} it is shown that the spin-lattice couplings $U^1_k$ and $U^2_{kk'}$ generate two relevant terms to explain the changes of sound velocity and attenuation in MnSc$_2$Se$_4$. The corresponding minimal Hamiltonian is then written as:
\begin{equation}\label{eq13}
    \begin{split}
        \mathcal{H} &= \mathcal{H}_{s} + \mathcal{H}_{l} + \mathcal{H}_{sl},\\
        \mathcal{H}_{l} &= \sum_{k\lambda} \omega_{k\lambda}^0 c_{k\lambda}^\dagger c_{k\lambda}, \\
        \mathcal{H}_{s} &= E_0 +\Delta E + \frac{1}{2}\sum_{q} \Psi_{q}^\dagger \bar{E}_{q} \Psi_q,\\
        \mathcal{H}_{sl} &=\frac{1}{2}\sum_k \Gamma_k^0 C_k C_{-k} + \sum_{qk} \Psi_{q+k}^\dagger \bar{M}_{kq} \Psi_q C_k,\\
    \end{split}{}
\end{equation}{}
where the vertices $\Gamma_k^0$ and $\bar{M}_{kq}$ are defined in Supplemental Material \cite{supp_mat}. We define the magnon Green function as $\bar{G}_0^s(q, \tau) = - \left\langle \mathcal{T}\Psi_q(\tau) \Psi_q(0)^\dagger \right\rangle$ where $\mathcal{T}$ is the chronological-order operator. Using standard functional-integral techniques \cite{altland2010condensed,saenger1995effect, woods2001magnon, cheng2008magnon}, the phonon self-energy is written as : 
\begin{equation}\label{eq21}
    \begin{split}
        \Sigma(k,i\nu_n) &= \Gamma_k^0 - \frac{1}{\beta N} \sum_{q, i\nu_m,i,j}\left(\bar{G}_0^s(q,i\nu_n)\right)_{ii}\left(\bar{M}_{q,q+k}\right)_{ij}\\ & \ \ \ \ \ \ \ \times\left(\bar{G}_0^s(q+k,i\nu_n+i\nu_m)\right)_{jj}\left(\bar{M}_{q+k,-k}\right)_{ji},
    \end{split}{}
\end{equation}{}
where $i\nu_n$ is a Matsubara frequency. These two interaction terms and their associated self-energy contributions are represented schematically in Fig. \hyperref[fig5]{\ref*{fig5}}. 
For the ultrasound results presented in Fig. \hyperref[fig2]{\ref*{fig2}}, we set an external perturbation in the form of a displacement wave $\mathbf{u}(\mathbf{r},t) = \mathbf{u}_0 \text{exp}(i(\mathbf{k}\cdot\mathbf{r}-\omega t))$, where $\mathbf{u}_0 \parallel \boldsymbol{\epsilon}_k$ is the sound-wave polarization. In our case, we have $\mathbf{k} \parallel [111]$. For the used frequencies below $f \approx 100$ MHz, we are in the limit $\mathbf{k}\cdot\boldsymbol{\delta} \ll 1$, where $\boldsymbol{\delta}$ is of the order of the interatomic distances \cite{luthi2007physical}. In this limit, the phonon dispersion along the $[111]$ axis is linear $\omega_k^0 = vk$. The sound-velocity change as a function of magnetic field corresponds then to the change of the phonon dispersion, which is evaluated up to a shift $\Delta_0$ as the real part of the phonon self-energy $\Sigma(k,i\nu_n)$: 
\begin{equation}\label{eq15}
    \begin{split}
   \frac{\Delta v}{v} &= \lim_{k\rightarrow 0} \frac{\Delta \omega_q^0}{\omega_k^0} = \lim_{k\rightarrow 0} \frac{\text{Re} \Sigma(k,i\nu_n)_{i\nu_n \rightarrow \omega_k^0 + i0^+}}{\omega_k^0} + \Delta_0\\
   &= \frac{\xi^2 S |J_1|}{v^2} \left[S \Sigma_1 + \Sigma_2\right] + \Delta_0,
    \end{split}{}
\end{equation}{}
where $\Sigma_1$, $\Sigma_2$ and $\Delta_0$ are dimensionless. The first term is a constant background [Fig. \hyperref[fig6]{\ref*{fig6}(a)}]:
\begin{equation}\label{eq16}
    \begin{split}
        &\Sigma_{1} = \frac{1}{2} \sum_{\delta_\alpha} \frac{J_{\delta_\alpha}}{|J_1|} \left(\sin\theta^2 + \cos\theta^2\cos(\mathbf{q}_0 \cdot \boldsymbol{\delta}_\alpha)\right)\\
        & \ \ \ \ \ \ \ \ \ \ \ \ \times\left(\frac{\mathbf{k} \cdot \boldsymbol{\delta}_\alpha}{||\mathbf{k}||}\right)^2(\boldsymbol{\delta}_\alpha \cdot \boldsymbol{\epsilon}_{k})^2.
    \end{split}{}
\end{equation}{}
This term leads to an increase of the sound velocity with increasing $H$ up to a saturation above $H_c$. The second term is evaluated at $T = 0$ and corresponds to a magnon-pair creation bubble:
\begin{equation}\label{eq17}
    \begin{split}
        &\Sigma_2 = -\frac{|J_1| S}{4N}\sum_q \left[ \frac{|\bar{M}_q^{23}|^2}{E_q^++E_{-q}^+} + \frac{|\bar{M}_q^{14}|^2}{E_q^-+E_{-q}^-} \right. \\
        & \ \ \ \ \ \ \ \ \ \ \ \ \ \ \ \ \ \ \ \ \ \ \ \ \ \ \ \ \ \   \left.+ \frac{|\bar{M}_q^{24}|^2+|\bar{M}_q^{13}|^2}{E_q^++E_{-q}^-}\right],
    \end{split}{}
\end{equation}{}
where $\bar{M}_{q} = \frac{2\sqrt{2}\sqrt{v}}{|J_1|\xi S} \frac{1}{k} \lim_{k\rightarrow 0} \bar{M}_{q,q+k}$. The sum is evaluated using standard Monte Carlo integration over the first Brillouin zone of the diamond lattice, and the result is shown in Fig. \hyperref[fig6]{\ref*{fig6}(b)}. 

Due to the presence of the magnon energies $E_q^\pm$ in the denominator of Eq. \hyperref[eq17]{(\ref*{eq17})}, the major contribution will be associated to the regions where $E_q^\pm$ is small, i.e., by the soft modes visualized in Fig. \hyperref[fig4]{\ref*{fig4}}. With increasing magnetic field, we identify two regimes. First, at low field the phonon self-energy increases with the magnetic field, for both phonon polarizations. This increase is related to the opening of the gap at the soft-mode position $\mathbf{q}_0$ [Fig. \hyperref[fig4]{\ref*{fig4}(b)}], which suppresses a contribution to $\Sigma_2$. At higher fields, we find a pronounced softening of the phonon self-energy for the longitudinal mode, leading to a minimum of $\Sigma_2$ at $H = 0.9 H_c$. Remarkably, for the transverse mode, this softening is absent.

This softening can be explained as follows: on the one hand, as $H$ increases, the magnon velocity $v_{mag}$ decreases, thus decreasing the value of the denominator in Eq. \hyperref[eq17]{(\ref*{eq17})} and increasing the magnitude of $\Sigma_2$. On  the other hand, the anomalous vertices $|\bar{M}_q^{23}|$, $|\bar{M}_q^{14}|$, $|\bar{M}_q^{24}|$ and $|\bar{M}_q^{23}|$ correspond to the expectation values of $a_{q}^\dagger a_{-q}^\dagger$, $b_{q}^\dagger b_{-q}^\dagger$, $a_{q}^\dagger b_{-q}^\dagger$, and $b_{q}^\dagger a_{-q}^\dagger$, respectively. These expectation values are equal to zero in the polarized state, and thus $\Sigma_2$ has to vanish for $H\rightarrow H_c$.

Thus, the phonon softening observed for the longitudinal polarization has a purely quantum origin: it arises from the quantum fluctuations of the antiferromagnetic state, which is not an exact ground state of the Heisenberg Hamiltonian. As the magnetic field is increased, the magnon spectrum softens, changing continuously from a linear dispersion, typical for antiferromagnetic systems, to a quadratic dispersion, typical for ferromagnetic systems. This induces a progressive increase of the phonon softening due to the interactions with spin-wave fluctuations. However, since the polarized state is the exact ground state of the Heisenberg Hamiltonian, the spin-wave fluctuations at zero temperature and the associated phonon softening are suppressed above the saturation field, leading to the characteristic feature shown in Fig. \hyperref[fig6]{\ref*{fig6}(b)}.

Furthermore, the magnon-pair creation bubble gives the only contribution to the ultrasound attenuation:
\begin{equation}\label{eq18}
    \begin{split}
        \Delta \alpha & = -\lim_{k\rightarrow 0} \frac{\text{Im}\Sigma(k,i\nu_n)_{i\nu_n \rightarrow \omega_k^0 + i\eta}}{\omega_k^0} = \frac{\pi\xi^2 |J_1|^2 S^2}{4Nv^2} \alpha_2 \\
        \alpha_2&= \sum_q \left[ |\bar{M}_q^{24}|^2 \delta_\eta(E_q^++E_{-q}^+)  + |\bar{M}_q^{13}|^2\delta_\eta(E_q^-+E_{-q}^-)   \right. \\
        & \ \ \ \ \ \ \ \ \ \ \ \ \ \ \ \ \ \ \ \ \ \ \ \left. + (|\bar{M}_q^{14}|^2+|\bar{M}_q^{23}|^2)\delta_\eta(E_q^++E_{-q}^-)\right]  \ ,
    \end{split}{}
\end{equation}{}
where $\delta_\eta(x) = \frac{1}{\pi}\eta/(\eta^2 + x^2)$ is the delta function for scattering time  $\eta = 10^{-4} |J_1| > 0$. The only constraint on $\eta$ is to be positive and much smaller than the other energy scales of the system, an for an arbitrary value of $\eta> 0$ Eq. \hyperref[eq18]{(\ref*{eq18})} fixes the value of the attenuation up to a constant factor. The result is shown in Fig. \hyperref[fig2]{\ref*{fig2}}, in good agreement with the experiment.

\subsection{Comparison to ultrasound experiments}

The theory reproduces qualitatively the observed features of the sound-velocity changes in MnSc$_2$Se$_4$. This is caused by a hardening of $\Sigma_1$ from zero field to the fully polarized state, which is present for both longitudinal and transverse acoustic modes, and a softening of $\Sigma_2$ in the intermediate-field regime, which is present only for the longitudinal acoustic mode.

The field dependent change of $\frac{\Delta v}{v}$ in Eq. \hyperref[eq15]{(\ref*{eq15})} depends only on two known parameters: the value of the spin $S = 5/2 $ and the saturation field  $\mu_0 H_c = 15.4$ T. The only free parameter of the theory is the magnetoelastic coupling $\xi$, which appears in the constant factor $ \frac{\xi^2 S |J_1|}{v^2}$.

For the longitudinal acoustic mode, the comparison between theory and experiments is shown in Fig. \hyperref[fig2]{\ref*{fig2}}. The constant was set to $ \frac{\xi^2 S |J_1|}{v^2} = 3 \times 10^{-4}$. The agreement is particularly good at high field. At low field, the theory predicts an increase of the sound velocity, associated to the suppression of the soft mode $\mathbf{q}_0$ [Fig. \hyperref[fig6]{\ref*{fig6}(b)}], which is not observed experimentally. Thus, the assumption of a spin-wave spectrum for MnSc$_2$Se$_4$ appears to be more justified above a certain threshold field. In Ref. \cite{lee2008theory}, it has been suggested that for MnSc$_2$S$_4$ at zero magnetic field, the spin spiral resides in some specific planes, due to the presence of magnetic anisotropy. This magnetic anisotropy has been proposed to originate either from dipolar interactions and covalence effects resulting in a spin density redistribution from Mn$^{2+}$ $d$ orbitals to the surrounding chalcogenide $p$ orbitals, or to spin-orbit effects \cite{lee2008theory}. Thus, the presence of a soft mode at $\mathbf{q}_0$ in a direction perpendicular to the magnetic field might be a wrong assumption, and including anisotropy effects would be a reasonable extension of this work.

For the transverse acoustic mode, the theory overestimate the magnitude of the hardening due to the factor $1/v^2$ present in Eq. \hyperref[eq15]{(\ref*{eq15})}. This effect might also be related to an oversimplification of the spin-lattice coupling, which takes into accounts only the distance between different Mn$^{2+}$ ions in the dependence of the exchange integrals in Eq. \hyperref[eq15]{(\ref*{eq15})}, and do not consider the effect of intermediate Se$^{2-}$ anions for example. Nevertheless, the qualitative features such as the absence of phonon softening is well reproduced by the model.

\section{Conclusion}

In this work, we have investigated the field-dependent magnetization, sound-velocity and attenuation changes in MnSc$_2$Se$_4$ single crystals at low temperatures. We performed magnetization measurements in pulsed fields up to 23 T, as well as ultrasound experiments in static fields up to 17 T. We also provide a minimal model that reproduces qualitatively our observations.

Our results show the presence of a saturated state above 15.4 T, where the magnetization reaches 4.75 $\mu_B$/f.u., close to the value $2S = 5$ $\mu_B$/f.u. expected for $S = 5/2$ Mn$^{2+}$ ions with quenched orbital moment. We have not observed any sharp transition in the magnetic and magnetoelastic properties between the zero-field state and the polarized state. However, we found a significant phonon softening for the longitudinal acoustic mode parallel to the field direction along[111], which is absent in the transverse acoustic mode.

The presented microscopic model is based on linear spin-wave theory, and supposes that the dynamics of MnSc$_2$Se$_4$ in magnetic field at low temperatures is dominated by spin-wave fluctuations. We can qualitatively reproduce the evolution of sound-velocity with respect to magnetic field with a Hamiltonian taking into account two terms arising from the exchange-striction mechanism: a constant background term and a quantum fluctuation term. The background term leads to a phonon hardening for both longitudinal and transverse polarizations. In contrast, the fluctuation term contributes significantly only for the longitudinal polarization and generates a softening at intermediate fields. Up to a constant factor, the model is in qualitative agreement with our experimental results without any free parameter.
Thus, we conclude that ultrasound measurements at low temperatures permits clearly to identify quantum fluctuation effects in frustrated magnets. Our study also shows how the use of a microscopic model permits to take into account the phonon polarization in a precise way by considering the real microscopic displacements of the Mn$^{2+}$ ions in the exact crystal structure. 

While this picture is simple enough to interpret the sound velocity changes in the high-field regime, there is a sensitive mismatch at low field.  This difference might be attributed to an oversimplification of the model, which neglect the role of the Sc$^{3+}$ and Se$^{2-}$ ions in the magnetic couplings, leading to possible anisotropy effects. A small magnetization anisotropy of 2\% between magnetic field parallel and perpendicular to [111] has been observed (Supplemental Material \cite{supp_mat}). 

Thus, a better characterization of the magnetic state at low magnetic field would give new insights to the magnetism of MnSc$_2$Se$_4$. In particular, the study of sound propagation $\mathbf{k}$ and magnetic field $\mathbf{H}$ in directions different than the [111] axis could give insights about possible anisotropy effects in this material.

\section{Aknowledgments}
We thank A. Hauspurg for his help during the ultrasound experiments, and O. Zaharko for fruitful discussions. We acknowledge the support of the High Magnetic Field Laboratory (HLD) at Helmholtz-Zentrum Dresden-Rossendorf (HZDR), a member of the European Magnetic Field Laboratory (EMFL), the Deutsche Forschungsgemeinschaft (DFG) through SFB 1143, and the Würzburg-Dresden Cluster of Excellence on Complexity and Topology in Quantum Matter ct.qmat (EXC 2147, Project No. 390858490). The support of  Project No. ANCD 20.80009.5007.19 (Moldova) is also acknowledged.

\bibliography{Biblio/biblio_JSourd_MNSc2Se42.bib}

\end{document}


\title{Supplemental Material to "Magnon-phonon interactions in the spinel compound MnSc$_2$Se$_4$"}

\author{ J. Sourd}
\affiliation{Hochfeld-Magnetlabor Dresden (HLD-EMFL) and Würzburg-Dresden Cluster of Excellence ct.qmat,
Helmholtz-Zentrum Dresden-Rossendorf, 01328 Dresden, Germany}
\author{Y. Skourski}
\affiliation{Hochfeld-Magnetlabor Dresden (HLD-EMFL) and Würzburg-Dresden Cluster of Excellence ct.qmat,
Helmholtz-Zentrum Dresden-Rossendorf, 01328 Dresden, Germany}
\author{L. Prodan}
\affiliation{Experimental Physics V, University of Augsburg, 86135 Augsburg, Germany}
\affiliation{Institute of Applied Physics, MD 2028, Chisinau, Republic of Moldova}
\author{V. Tsurkan}
\affiliation{Experimental Physics V, University of Augsburg, 86135 Augsburg, Germany}
\affiliation{Institute of Applied Physics, MD 2028, Chisinau, Republic of Moldova}
\author{A. Miyata}
\affiliation{Institute for Solid State Physics, The University of Tokyo, Kashiwa, Chiba 277-8581, Japan}
\author{J. Wosnitza}
\affiliation{Hochfeld-Magnetlabor Dresden (HLD-EMFL) and Würzburg-Dresden Cluster of Excellence ct.qmat,
 Helmholtz-Zentrum Dresden-Rossendorf, 01328 Dresden, Germany}
\affiliation{Institut für Festkörper- und Materialphysik, TU Dresden, 01062 Dresden, Germany}
\author{S. Zherlitsyn}
\affiliation{Hochfeld-Magnetlabor Dresden (HLD-EMFL) and Würzburg-Dresden Cluster of Excellence ct.qmat,
Helmholtz-Zentrum Dresden-Rossendorf, 01328 Dresden, Germany}

\maketitle

\onecolumngrid

\section{Composition analysis}
\begin{figure}[h!]
    \centering
    \includegraphics[width=.9\linewidth]{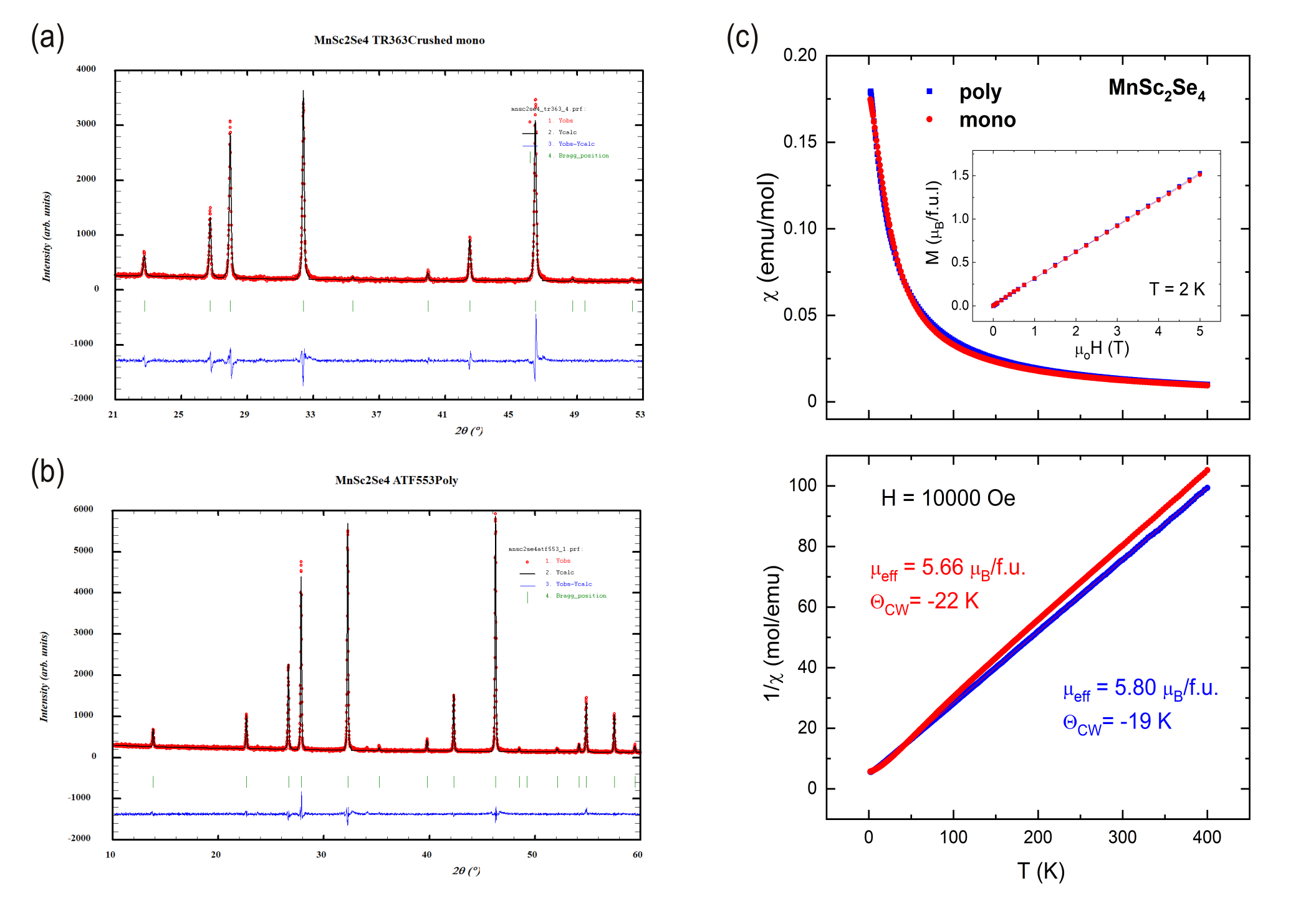}
    \caption{(a) Powder x-ray diffraction (PXRD) pattern of a crushed MnSc2Se4 single crystal, refined with the Rietveld method. The refinement parameters are: Bragg R-factor = 12.8, Rf = 15.0, Chi2 = 6.20, a0 = 11.0451 (5), x = 0.25689 (17). (b) PXRD pattern of a polycrystalline MnSc2Se4 sample, refined with the Rietveld method. The  refinement parameters are: Bragg R-factor = 3.54, Rf = 4.40, Chi2 = 4.30, a0 = 11.0887 (1), x = 0.25687 (5). (c)  DC magnetic susceptibility vs. temperature of single crystalline and polycrystalline MnSc$_2$Se$_4$ samples.}\label{figS0}
\end{figure}

We characterized a crushed single crystal and a polycrystalline sample by powder x-ray diffraction (PXRD) (Fig. \hyperref[figS0]{\ref*{figS0}(a)} and Fig. \hyperref[figS0]{\ref*{figS0} (b)}).  The refinement of the PXRD pattern for the crushed single crystal is good taking into account the small sample mass (Bragg R-factor = 12.8, Rf = 15.0, Chi2 = 6.20, a0 = 11.0451(5), x = 0.25689(17)) and shows the absence of a secondary phase. The analysis of the polycrystalline sample resulted in (Bragg R-factor = 3.54, Rf = 4.40, Chi2 = 4.30, a0 = 11.0887 (1), x = 0.25687 (5)), again with no indication of a secondary phase. We measured the magnetization of a single-crystalline and polycrystalline sample (Fig. \hyperref[figS0]{\ref*{figS0} (c)}), which shows good agreement.

\section{Diagonalization of the spin-wave Hamiltonian}\label{Colpa_algo}

In this Appendix we present the algorithm of Colpa \cite{colpa1978diagonalization} for a general Bogoliubov transformation. We write the spin-wave Hamiltonian of Eq. (9) in matrix form, with $\Phi_q^\dagger = (a_{q}^\dagger, b_{q}^\dagger, a_{-q}, b_{-q})$:
\begin{equation}\label{A1}
    \begin{split}
        \mathcal{H}_{s} &= E_0 + \frac{1}{2}\sum_{q}\left[ \Phi_q^\dagger \mathcal{H}_q  \Phi_q - 2A_{-q}\right],  \\
        \mathcal{H}_q &= \begin{pmatrix} A_{q} & C_{q} & B_{q} & D_{q} \\ C_{q}^* & A_{q} & D_{q}^* & B_{q} \\ B_{q} & D_{q} & A_{-q} & C_{-q}^* \\ D_{q}^* & B_{q} & C_{-q} & A_{-q} \end{pmatrix}.
    \end{split}{}
\end{equation}{}
This Hamiltonian has to be diagonalized with the constraint of preserving the boson-commutation relation: 
\begin{equation}\label{A2}
    \begin{split}
        \left[\Phi_q^\dagger, \Phi_q\right] = \bar{\mathcal{J}} = \begin{pmatrix} 1 & 0 & 0 & 0 \\ 0 & 1 & 0 & 0 \\ 0 & 0 & -1 & 0 \\ 0 & 0 & 0 & -1 \end{pmatrix}.
    \end{split}{}
\end{equation}{}
As a first step, we use the Cholesky decomposition in order to write $ \mathcal{H}_q = K^\dagger K$. Then, we diagonalize the matrix $K\bar{\mathcal{J}}K^\dagger$, and write it in a diagonal form as $L = U^\dagger K\bar{\mathcal{J}}K^\dagger U$. The eigenvalues of $L$ then correspond to plus and minus the magnon energies $\bar{E}$. Sorting the eigenvalues of $L$ in ascending order with $E_{q}^- > E_q^+ > -E_{-q}^+ > -E_{-q}^-$, the associated eigenvectors of $L$ correspond to $\Psi_q^\dagger = (\beta_q^\dagger, \alpha_q^\dagger , \alpha_{-q},  \beta_{-q})$, such that:  
\begin{equation}\label{A3}
    \begin{split}
        &\bar{E}_q  = \bar{\mathcal{J}} L = \bar{R}_q^\dagger \mathcal{H}_q \bar{R}_q, \ \ \Phi_q  = \bar{R}_q \Psi_q,  \ \ \bar{R}_q = K^{-1}U\sqrt{\bar{E}},\\
        &\Rightarrow \mathcal{H}_{s} = E_0 +\Delta E + \sum_{q} \left(E^+_q \alpha_q^\dagger \alpha_q + E^-_q \beta_q^\dagger \beta_q \right),\\
      &\Delta E  = - \sum_q A_q + \frac{1}{2} \sum_q \left( E_q^+ + E_q^-\right).
    \end{split}{}
\end{equation}{}
Now, we consider the magnon dynamics in the limit $q\rightarrow0$. In this limit, we can neglect the imaginary parts of $C_q$ and $D_q$, and the $\sin\theta\sin(\mathbf{q}_0 \cdot \boldsymbol{\delta}_{2})$ term in $A_q$, so that $A_q = A_{-q}$. This permits us to write the explicit Bogoliubov transformation as in \cite{rastelli1990frustration}:
\begin{equation}\label{A4}
    \begin{split}
        \alpha_q &= u_q (a_q + b_q) + v_q (a_{-q}^\dagger + b_{-q}^\dagger),   \\
        \beta_q  &=  r_q (a_q-b_q) + t_q (a_{-q}^\dagger - b_{-q}^\dagger),  \\
        u_q & = \sqrt{\frac{E_q^+ + A_q + C_q}{4E_q^+}}, \ 
        v_q  = \sqrt{\frac{-E_q^+ + A_q + C_q}{4E_q^+}},\\
        r_q & = \sqrt{\frac{E_q^- + A_q - C_q}{4E_q^-}}, \
        t_q  = \sqrt{\frac{-E_q^- + A_q - C_q}{4E_q^-}},\\
    \end{split}{}
\end{equation}{}
where $E_q^\pm$ are given by:
\begin{equation}\label{A5}
    \begin{split}
        E^\pm_q  &= \sqrt{\left(A_q \pm C_q\right)^2 - \left(B_q \pm D_q\right)^2}.\\
    \end{split}{}
\end{equation}{}
From these expressions, we can evaluate the magnon velocity along $[111]$ as $v_{mag} = \lim_{q\rightarrow 0} \frac{\partial E_q^+}{\partial q }$. In the case of MnSc$_2$Se$_4$, we have: $a_0 = 11.1$ \r{A} for the lattice constant, $|J_1| = 0.33$ K and $S= 5/2$, which gives:
\begin{equation}\label{A6}
    \begin{split}
        &v_{mag} =\lim_{q\rightarrow 0} \frac{\partial E_q^+}{\partial q } =\frac{a k_B}{\hbar}\frac{S|J_1|}{\sqrt{2}} \left[\sum_{\delta_\alpha \delta_{\alpha'}}p_{\delta_\alpha}p_{\delta_{\alpha'}}  \cos(\mathbf{q}_0 \cdot \boldsymbol{\delta}_\alpha)\left[(1-\sin\theta^2) \cos(\mathbf{q}_0 \cdot \boldsymbol{\delta}_{\alpha'})-\cos\theta^2\right]\left(\frac{\mathbf{q} \cdot \boldsymbol{\delta}_{\alpha'}}{||\mathbf{q}||}\right)^2\right]^{1/2} \\
        &\Rightarrow v_{mag} \leq v_{mag}(H = 0) \approx 470 \  \text{m}/\text{s}, 
    \end{split}{}
\end{equation}{}
where $p_{\delta_\alpha} = J_{\delta_\alpha}/|J_1|$ are the dimensionless couplings constants and $\delta_{\alpha}, \delta_{\alpha'}$ run over $\delta_1$, $\delta_2$ and $\delta_3$. The magnon velocity is, thus, approximately one order of magnitude smaller than the phonon velocity measured experimentally (Table I). Furthermore, we remark that the magnon velocity vanishes in the limit $H \rightarrow H_c \Leftrightarrow \sin \theta \rightarrow 1$.
\section{Magnetization}\label{M_and_chi}
In this section, we evaluate the magnetization and the magnetization derivative as derivatives of the ground-state energy with respect to the magnetic field \cite{zhitomirsky1998magnetization}. Since we do not have access to analytical expressions for the eigenvalues $\bar{E}_q$, we will use perturbation theory in order to evaluate the derivative as $\frac{\partial E}{\partial h} = \lim_{\delta h\rightarrow 0} \frac{E(h+\delta h)-E(h)}{\delta h}$. We write $\mathcal{H} = \mathcal{H}_s + \mathcal{V}$, where $\mathcal{V} =  - \sum_i \delta \mathbf{h}\cdot \left(\mathbf{S}_i^a+\mathbf{S}_i^b\right)$ is the perturbation in external magnetic field. Using $ \delta \mathbf{h}\cdot \mathbf{S}_i^\alpha = \sin \theta \delta h S_i^{\alpha \parallel}$, one need to the evaluate the average values of $a_i^\dagger a_i$ and  $b_i^\dagger b_i$. With $\sin\theta = H/H_c$, we obtain:
\begin{equation}\label{B1}
    \begin{split}
        &m =  \frac{SH}{H_c}\left[1 + \frac{1}{2S} - \frac{1}{4SN}\sum_q \Tr \bar{R}_q^\dagger \bar{D} \bar{R}_q \right].
    \end{split}
\end{equation}
where $\bar{D} = \left\langle \Psi_q^\dagger \Psi_q \right\rangle$ is the average value of Bogoliubov rotated operators. The first term corresponds to the classical magnetization of a Heisenberg antiferromagnet, and the second and third terms correspond to the field-dependent quantum correction to the Mn$^{2+}$ magnetic moment, arising from the zero-point motion of spin waves. These terms give rise to an upturn of $\partial M/\partial H$ which should be small for large values of $S$, and which is not observed experimentally.

\section{Exchange striction terms}\label{matrices_MN}

Replacing the spin operators by their Holstein-Primakoff expansion in the $U_k^1$ and $U_{kk'}^2$ terms of Eq. (11), and using the Bogoliubov rotated basis for the magnon operators $\Psi_q^\dagger = (\beta_q^\dagger, \alpha_q^\dagger , \alpha_{-q},  \beta_{-q})$ we obtain four different terms:
\begin{equation}\label{eq14}
    \begin{split}
        \mathcal{H} &= \mathcal{H}_{s} + \mathcal{H}_{l} + \mathcal{H}_{sl}^1 + \mathcal{H}_{sl}^2 + \mathcal{H}_{sl}^3 + \mathcal{H}_{sl}^4 \\
        \mathcal{H}_{l} &= \sum_{k\lambda} \omega_{k\lambda}^0 c_{k\lambda}^\dagger c_{k\lambda} \\
        \mathcal{H}_{s} &= E_0 +\Delta E + \frac{1}{2}\sum_{q} \Psi_{q}^\dagger \bar{E}_{q} \Psi_q\\
        \mathcal{H}_{sl}^1 &=\frac{1}{2}\sum_k \Gamma_k^0 C_k C_{-k} \\
        \mathcal{H}_{sl}^2 &= \sum_{qk} \Psi_{q+k}^\dagger \bar{M}_{kq} \Psi_q C_k\\
         \mathcal{H}_{sl}^3 &= \sum_k \Psi_k^\dagger \cdot \boldsymbol{\gamma}_k C_k \\
        \mathcal{H}_{sl}^4 &= \frac{1}{2}\sum_{kk'q}\left[\Psi_{k+k'+q} ^\dagger \bar{N}_{kk'q} \Psi_q - \Delta A_{kq}\delta_{k+k'}\right] C_kC_{k'}. 
    \end{split}{}
\end{equation}{}
The first contribution is a constant background:
\begin{equation}\label{C1}
    \begin{split}
        \mathcal{H}_{sl}^1 &=\frac{1}{2}\sum_k \Gamma_k^0 C_k C_{-k}, \\ 
        \Gamma_k^0 &= \frac{\xi^2 S^2}{\omega_k^0}\sum_{\delta_\alpha} J_{\delta_\alpha} \left(\sin\theta^2 + \cos\theta^2\cos(\mathbf{q}_0 \cdot \boldsymbol{\delta}_\alpha)\right)\left(1-\cos(\mathbf{k} \cdot \boldsymbol{\delta}_\alpha)\right)(\boldsymbol{\delta}_\alpha \cdot \boldsymbol{\epsilon}_{k})^2.
    \end{split}{}
\end{equation}{}
The second contribution is a two-magnons one-phonon interaction:
\begin{equation}\label{C3}
    \begin{split}
        \mathcal{H}_{sl}^2 &= \sum_{kq} \Psi_{q+k}^\dagger \bar{M}_{qk}  \Psi_q  C_k, \ \ \ \ \bar{M}_{qk} = -\frac{i\xi S}{2\sqrt{2\omega_k^0}}  (\bar{R}_{q+k})^\dagger \begin{pmatrix} A_{qk} & C_{qk} & B_{qk} & D_{qk} \\ C_{(q+k)k}^* & A_{qk} & D_{(q+k)k}^* & B_{(q+k)k}^* \\ B_{qk} & D_{qk} & A_{-q-k} & C_{-q-k}^* \\ D_{(q+k)k}^* & B_{(q+k)k}^* & C_{(-q-k)-k} & A_{-q-k} \end{pmatrix} \bar{R}_{q} \\
        A_{qk} &= -\sum_{\delta_\alpha}J_{\delta_\alpha}\left[\cos\theta^2\cos(\mathbf{q}_0 \cdot \boldsymbol{\delta}_\alpha) + \sin\theta^2\right]\left(1-e^{i\mathbf{k} \cdot \boldsymbol{\delta}_\alpha}\right)(\boldsymbol{\delta}_\alpha \cdot \boldsymbol{\epsilon}_{k})\\
        &+ \frac{J_2}{4} \sum_{\delta_2} \left[ \left(1 + \sin\theta^2 \right)\cos(\mathbf{q}_0 \cdot \boldsymbol{\delta}_2) + \cos\theta^2 +2i\sin\theta\sin(\mathbf{q}_0 \cdot \boldsymbol{\delta}_{2}) \right] \left(e^{i\mathbf{q} \cdot \boldsymbol{\delta}_2}+e^{-i(\mathbf{q}+\mathbf{k}) \cdot \boldsymbol{\delta}_2} \right)\left(1-e^{i\mathbf{k} \cdot \boldsymbol{\delta}_2}\right)(\boldsymbol{\delta}_2 \cdot \boldsymbol{\epsilon}_{k}) \\
        B_{qk} &= \frac{J_2}{2} \sum_{\delta_2} \cos\theta^2\left[\cos(\mathbf{q}_0 \cdot \boldsymbol{\delta}_2) - 1  \right]e^{i\mathbf{q} \cdot \boldsymbol{\delta}_2}\left(1-e^{i\mathbf{k} \cdot \boldsymbol{\delta}_2}\right)(\boldsymbol{\delta}_2 \cdot \boldsymbol{\epsilon}_{k})\\
        C_{qk} &= \frac{1}{2}  \sum_{\delta_{\alpha'} = \delta_1, \delta_3} J_{\delta_{\alpha'}} \left[\left(1 + \sin\theta^2 \right)\cos(\mathbf{q}_0 \cdot \boldsymbol{\delta}_{\alpha'})+\cos\theta^2 +2i\sin\theta\sin(\mathbf{q}_0 \cdot \boldsymbol{\delta}_{\alpha'}) \right]e^{i\mathbf{q} \cdot \boldsymbol{\delta}_2}\left(1-e^{i\mathbf{k} \cdot \boldsymbol{\delta}_{\alpha'}}\right)(\boldsymbol{\delta}_{\alpha'} \cdot \boldsymbol{\epsilon}_{k})\\
        D_{qk} &= \frac{1}{2}  \sum_{\delta_{\alpha'} = \delta_1, \delta_3} J_{\delta_{\alpha'}}  \cos\theta^2\left[\cos(\mathbf{q}_0 \cdot \boldsymbol{\delta}_{\alpha'}) - 1  \right]e^{i\mathbf{q} \cdot \boldsymbol{\delta}_{\alpha'}}\left(1-e^{i\mathbf{k} \cdot \boldsymbol{\delta}_{\alpha'}}\right)(\boldsymbol{\delta}_{\alpha'} \cdot \boldsymbol{\epsilon}_{k}),
    \end{split}{}
\end{equation}{}
where the matrix $\bar{R}_k$, defined in section \ref{Colpa_algo}, is the transformation matrix between the Holstein-Primakoff bosons $a, b$ and the Bogoliubov rotated bosons $\alpha, \beta$. The third contribution is a magnon-phonon hybridization term:
\begin{equation}\label{C2}
    \begin{split}
        \mathcal{H}_{sl}^3 &= \sum_k \Psi_k^\dagger \cdot \boldsymbol{\gamma}_k C_k \\ 
        \boldsymbol{\gamma}_k &= -\frac{i\xi S^{3/2}\cos\theta}{2\sqrt{\omega_k^0}} \bar{R}_k^\dagger \begin{pmatrix} \gamma_k^+ \\ \gamma_k^+ \\ \gamma_k^- \\ \gamma_k^- \end{pmatrix} \\
        \gamma_k^\pm &= \gamma_k^1 \pm \gamma_k^2\\
        \gamma_k^1 &= \sum_{\delta_\alpha}J_{\delta_\alpha}  (\boldsymbol{\delta}_\alpha \cdot \boldsymbol{\epsilon}_{k})\sin(\mathbf{q}_0 \cdot \boldsymbol{\delta}_\alpha) \left[\cos(\mathbf{k} \cdot \boldsymbol{\delta}_\alpha) -1\right]\\  
        \gamma_k^2 &= \sin\theta\sum_{\delta_\alpha}J_{\delta_\alpha}  (\boldsymbol{\delta}_\alpha \cdot \boldsymbol{\epsilon}_{k})\left(\cos(\mathbf{q}_0 \cdot \boldsymbol{\delta}_\alpha) - 1 \right) \sin(\mathbf{k} \cdot \boldsymbol{\delta}_\alpha),  
    \end{split}{}
\end{equation}{}
The fourth term is a two-magnons two-phonons interaction:
\small{
\begin{equation}\label{C4}
    \begin{split}
        \mathcal{H}_{sl}^4 &= \frac{1}{2}\sum_{qkk'}\left[ \Psi_{q+k+k'}^\dagger \bar{N}_{kk'q}  \Psi_q - \Delta A_{qk}\delta_{k+k'} \right] C_kC_k' 
        \\  \bar{N}_{kk'q} &= -\frac{\xi^2 S}{4N \sqrt{\omega_k^0\omega_{k'}^0}}  (\bar{R}_{q+k+k'})^\dagger \begin{pmatrix} A_{qkk'} & C_{qkk'} & B_{qkk'} & D_{qkk'} \\ C_{(q+k+k')kk'}^* & A_{qkk'} & D_{(q+k+k')kk'}^* & B_{(q+k+k')kk'}^* \\ B_{qkk'} & D_{qkk'} & A_{-q-k-k'} & C_{-q-k-k'}^* \\ D_{(q+k+k')kk'}^* & B_{(q+k+k')kk'}^* & C_{(-q-k-k')-k-k'} & A_{-q-k-k'} \end{pmatrix} \bar{R}_{q},
    \end{split}{}
\end{equation}{}}
\small{
\begin{equation}\label{C5}
    \begin{split}
        A_{qkk'} &= -\sum_{\delta_\alpha}J_{\delta_\alpha}\left[\cos\theta^2\cos(\mathbf{q}_0 \cdot \boldsymbol{\delta}_\alpha) + \sin\theta^2\right]\left(1-e^{i\mathbf{k} \cdot \boldsymbol{\delta}_\alpha}\right)\left(1-e^{i\mathbf{k'} \cdot \boldsymbol{\delta}_\alpha}\right)(\boldsymbol{\delta}_\alpha \cdot \boldsymbol{\epsilon}_{k})(\boldsymbol{\delta}_\alpha \cdot \boldsymbol{\epsilon}_{k'})\\
        &+ \frac{J_2}{4} \sum_{\delta_2} \left[ \left(1 + \sin\theta^2 \right)\cos(\mathbf{q}_0 \cdot \boldsymbol{\delta}_2) + \cos\theta^2 +2i\sin\theta\sin(\mathbf{q}_0 \cdot \boldsymbol{\delta}_{2}) \right] \left(e^{i\mathbf{q} \cdot \boldsymbol{\delta}_2}+e^{-i(\mathbf{q}+\mathbf{k}+\mathbf{k}') \cdot \boldsymbol{\delta}_2} \right)\left(1-e^{i\mathbf{k} \cdot \boldsymbol{\delta}_2}\right)\left(1-e^{i\mathbf{k'} \cdot \boldsymbol{\delta}_2}\right)(\boldsymbol{\delta}_2 \cdot \boldsymbol{\epsilon}_{k})(\boldsymbol{\delta}_2 \cdot \boldsymbol{\epsilon}_{k'}) \\
        B_{qkk'} &= \frac{J_2}{2} \sum_{\delta_2} \cos\theta^2\left[\cos(\mathbf{q}_0 \cdot \boldsymbol{\delta}_2) - 1  \right]e^{i\mathbf{q} \cdot \boldsymbol{\delta}_2}\left(1-e^{i\mathbf{k} \cdot \boldsymbol{\delta}_2}\right)\left(1-e^{i\mathbf{k'} \cdot \boldsymbol{\delta}_2}\right)(\boldsymbol{\delta}_2 \cdot \boldsymbol{\epsilon}_{k})(\boldsymbol{\delta}_2 \cdot \boldsymbol{\epsilon}_{k'})\\
        C_{qkk'} &= \frac{1}{2}\sum_{\delta_{\alpha'} = \delta_1, \delta_3} J_{\alpha'} \left[\left(1 + \sin\theta^2 \right)\cos(\mathbf{q}_0 \cdot \boldsymbol{\delta}_{\alpha'})+\cos\theta^2 +2i\sin\theta\sin(\mathbf{q}_0 \cdot \boldsymbol{\delta}_{\alpha'}) \right]e^{i\mathbf{q} \cdot \boldsymbol{\delta}_{\alpha'}}\\ 
        & \ \ \ \ \ \ \ \ \ \ \ \ \ \times\left(1-e^{i\mathbf{k} \cdot \boldsymbol{\delta}_{\alpha'}}\right)\left(1-e^{i\mathbf{k'} \cdot \boldsymbol{\delta}_{\alpha'}}\right)(\boldsymbol{\delta}_{\alpha'} \cdot \boldsymbol{\epsilon}_{k})(\boldsymbol{\delta}_{\alpha'} \cdot \boldsymbol{\epsilon}_{k'})\\
        D_{qkk'} &= \frac{1}{2}\sum_{\delta_{\alpha'} = \delta_1, \delta_3} J_{\alpha'}  \cos\theta^2\left[\cos(\mathbf{q}_0 \cdot \boldsymbol{\delta}_{\alpha'}) - 1  \right]e^{i\mathbf{q} \cdot \boldsymbol{\delta}_{\alpha'}}\left(1-e^{i\mathbf{k} \cdot \boldsymbol{\delta}_{\alpha'}}\right)\left(1-e^{i\mathbf{k'} \cdot \boldsymbol{\delta}_{\alpha'}}\right)(\boldsymbol{\delta}_{\alpha'} \cdot \boldsymbol{\epsilon}_{k})(\boldsymbol{\delta}_{\alpha'} \cdot \boldsymbol{\epsilon}_{k'})\\
        \Delta A_{qk} &= -\frac{\xi^2 S}{4N \omega_k^0} \left[\sum_{\delta_\alpha}J_{\delta_\alpha}\left[\cos\theta^2\cos(\mathbf{q}_0 \cdot \boldsymbol{\delta}_\alpha) + \sin\theta^2\right]\left(1-e^{i\mathbf{k} \cdot \boldsymbol{\delta}_\alpha}\right)\left(1-e^{-i\mathbf{k} \cdot \boldsymbol{\delta}_\alpha}\right)(\boldsymbol{\delta}_\alpha \cdot \boldsymbol{\epsilon}_{k})^2 \right.\\
        &\left. - \frac{J_2}{4} \sum_{\delta_2} \left[ \left(1 + \sin\theta^2 \right)\cos(\mathbf{q}_0 \cdot \boldsymbol{\delta}_2) + \cos\theta^2 +2i\sin\theta\sin(\mathbf{q}_0 \cdot \boldsymbol{\delta}_{2}) \right] \left(e^{i\mathbf{q} \cdot \boldsymbol{\delta}_2}-e^{-i(\mathbf{q}+\mathbf{k}+\mathbf{k}') \cdot \boldsymbol{\delta}_2} \right)\left(1-e^{i\mathbf{k} \cdot \boldsymbol{\delta}_2}\right)\left(1-e^{-i\mathbf{k} \cdot \boldsymbol{\delta}_2}\right)(\boldsymbol{\delta}_2 \cdot \boldsymbol{\epsilon}_{k})^2 \right].
    \end{split}{}
\end{equation}{}}

\section{Self energies and $k\rightarrow 0$ limit}\label{selfE_et_k0}

The phonon self-energy is obtained using standard path-integral techniques. The contribution from the constant background is evaluated directly:
\begin{equation}\label{D1}
    \begin{split}
        \Sigma^1(k) = \Gamma_k^0
    \end{split}{}
\end{equation}{}
We define the magnon Green function as $\bar{G}_0^s(k, \tau) = - \left\langle \mathcal{T}\Psi_k(\tau) \Psi_k^\dagger(0) \right\rangle$ where $\mathcal{T}$ is the chronological order operator. This gives after Fourier transform $\bar{G}_0^s(k, i\nu_n) = (i\nu_n \bar{\mathcal{J}} - \bar{E}_k)^{-1}$ where $\bar{\mathcal{J}}_{ij} = \left(\delta_{i,1}\delta_{j,1} + \delta_{i,2}\delta_{j,2}\right) - \left(\delta_{i,3}\delta_{j,3} + \delta_{i,4}\delta_{j,4}\right)$, and $\bar{E}_{kij} = E_k^-\delta_{i,1}\delta_{j,1} + E_k^+\delta_{i,2}\delta_{j,2} - E_{-k}^+\delta_{i,3}\delta_{j,3} -E_{-k}^- \delta_{i,4}\delta_{j,4}$. 

\

The phonon self-energy arising from the magnon-pair creation bubble is evaluated as:
\begin{equation}\label{D5}
    \begin{split}
        \Sigma^2(k,i\nu_n) &= - \frac{1}{\beta} \sum_{q, i\nu_m,i,j}\left(\bar{G}_0^s(q,i\nu_n)\right)_{ii}\left(\bar{M}_{q,q+k}\right)_{ij}\left(\bar{G}_0^s(q+k,i\nu_n+i\nu_m)\right)_{jj}\left(\bar{M}_{q+k,-k}\right)_{ji}.
    \end{split}{}
\end{equation}{}
In the $T\rightarrow0$ limit, only the terms with anomalous vertices will contribute and we have:
\begin{equation}\label{D6}
    \begin{split}
        \Sigma^2(k,i\nu_n) &=  \sum_{q}\frac{\left(\bar{M}_{q,k}\right)_{32}\left(\bar{M}_{q+k,-k}\right)_{23}}{i\nu_n + E_{-q}^+ + E_{q+k}^+} - \sum_{q} \frac{\left(\bar{M}_{q,k}\right)_{23}\left(\bar{M}_{q+k,-k}\right)_{32}}{i\nu_n - E_{q}^+ - E_{-q-k}^+}
        + \sum_q\frac{\left(\bar{M}_{q,k}\right)_{41}\left(\bar{M}_{q+k,-k}\right)_{14}}{i\nu_n + E_{-q}^- + E_{q+k}^-} - \sum_q\frac{\left(\bar{M}_{q,k}\right)_{14}\left(\bar{M}_{q+k,-k}\right)_{41}}{i\nu_n - E_{q}^- - E_{-q-k}^-} \\
        &+\sum_q \frac{ \left(\bar{M}_{q,k}\right)_{42}\left(\bar{M}_{q+k,-k}\right)_{24} }{i\nu_n + E_{-q}^- + E_{q+k}^+} -\sum_q \frac{ \left(\bar{M}_{q,k}\right)_{24}\left(\bar{M}_{q+k,-k}\right)_{42} }{i\nu_n - E_{q}^- - E_{-q-k}^+}
        +\sum_q \frac{\left(\bar{M}_{q,k}\right)_{31}\left(\bar{M}_{q+k,-k}\right)_{13}}{i\nu_n + E_{-q}^+ + E_{q+k}^-} -\sum_q \frac{ \left(\bar{M}_{q,k}\right)_{13}\left(\bar{M}_{q+k,-k}\right)_{31}}{i\nu_n - E_{q}^+ - E_{-q-k}^-}.
    \end{split}{}
\end{equation}{}
In the $k \rightarrow 0$ limit, we can replace $\left(1-e^{i\mathbf{k} \cdot \boldsymbol{\delta}_\alpha}\right)$ by $-i\left(\mathbf{k} \cdot \boldsymbol{\delta}_\alpha\right)$. We thus have $\bar{M}_{q+k,-k} = -\bar{M}_{q,k}$ and by taking out the prefactors we have:
\begin{equation}\label{D7}
    \begin{split}
        &\lim_{k\rightarrow 0} \frac{\text{Re}\Sigma^2(k,i\nu_n)_{i\nu_n \rightarrow \omega_k^0 + i0^+}}{\omega_k^0} = -\frac{\xi^2 |J_1|S}{v^2} \times \frac{|J_1| S}{4 N}\sum_q \left[ \frac{|\bar{M}_q^{23}|^2}{E_q^++E_{-q}^+} + \frac{|\bar{M}_q^{14}|^2}{E_q^-+E_{-q}^-} + \frac{|\bar{M}_q^{24}|^2+|\bar{M}_q^{13}|^2}{E_q^++E_{-q}^-}\right],\\
        &-\lim_{k\rightarrow 0} \frac{\text{Im}\Sigma^2(k,i\nu_n)_{i\nu_n \rightarrow \omega_k^0 + i\eta}}{\omega_k^0} = \frac{\pi\xi^2|J_1|^2S^2}{4Nv^2} \sum_q \left[ |\bar{M}_q^{24}|^2 \delta_\eta(E_q^++E_{-q}^+)  + |\bar{M}_q^{13}|^2\delta_\eta(E_q^-+E_{-q}^-)  + (|\bar{M}_q^{14}|^2+|\bar{M}_q^{23}|^2)\delta_\eta(E_q^++E_{-q}^-)\right],
    \end{split}{}
\end{equation}{}
where $\delta_\eta(x) = \frac{1}{\pi}\eta/(\eta^2 + x^2)$ is the delta function for scattering time $\eta$ and the matrix $\bar{M}_{q}$ is given by:
\begin{equation}\label{D8}
    \begin{split}
        \bar{M}_{q} &= \left(\bar{R}_{q}\right)^\dagger \begin{pmatrix} A_{q}' & C_{q}' & B_{q}' & D_{q}' \\ (C_{q}')^* & A_{q}' & (D_{q}')^* & (B_{q}')^* \\ B_{q}' & D_{q}' & A_{-q}' & (C_{-q}')^* \\ (D_{q}')^* & (B_{q}')^* & C_{-q}' & A_{-q}' \end{pmatrix} \bar{R}_{q} \\
    A_{q}' &= \sum_{\delta_\alpha}p_{\delta_\alpha}\left[\cos\theta^2\cos(\mathbf{q}_0 \cdot \boldsymbol{\delta}_\alpha) + \sin\theta^2\right]\left(\frac{\mathbf{k} \cdot \boldsymbol{\delta}_\alpha}{||\mathbf{k}||}\right)(\boldsymbol{\delta}_\alpha \cdot \boldsymbol{\epsilon}_{k})\\
    &- \frac{p_2}{2} \sum_{\delta_2} \left[ \left(1 + \sin\theta^2 \right)\cos(\mathbf{q}_0 \cdot \boldsymbol{\delta}_2) + \cos\theta^2 +2i\sin\theta\sin(\mathbf{q}_0 \cdot \boldsymbol{\delta}_{2})\right] e^(i\mathbf{q} \cdot \boldsymbol{\delta}_2)\left(\frac{\mathbf{k} \cdot \boldsymbol{\delta}_2}{||\mathbf{k}||}\right)(\boldsymbol{\delta}_2 \cdot \boldsymbol{\epsilon}_{k}) \\
    B_{q}' &= -\frac{p_2}{2} \sum_{\delta_2} \cos\theta^2\left[\cos(\mathbf{q}_0 \cdot \boldsymbol{\delta}_2) - 1  \right]e^{i\mathbf{q} \cdot \boldsymbol{\delta}_2}\left(\frac{\mathbf{k} \cdot \boldsymbol{\delta}_2}{||\mathbf{k}||}\right)(\boldsymbol{\delta}_2 \cdot \boldsymbol{\epsilon}_{k})\\
    C_{q}' &= -\frac{1}{2}\sum_{\delta_{\alpha'} = \delta_1, \delta_3} p_{\alpha'} \left[\left(1 + \sin\theta^2 \right)\cos(\mathbf{q}_0 \cdot \boldsymbol{\delta}_{\alpha'})+\cos\theta^2 +2i\sin\theta\sin(\mathbf{q}_0 \cdot \boldsymbol{\delta}_{\alpha'}) \right]e^{i\mathbf{q} \cdot \boldsymbol{\delta}_{\alpha'}}\left(\frac{\mathbf{k} \cdot \boldsymbol{\delta}_{\alpha'}}{||\mathbf{k}||}\right)(\boldsymbol{\delta}_{\alpha'} \cdot \boldsymbol{\epsilon}_{k})\\
    D_{q}' &= -\frac{1}{2}\sum_{\delta_{\alpha'} = \delta_1, \delta_3} p_{\alpha'}  \cos\theta^2\left[\cos(\mathbf{q}_0 \cdot \boldsymbol{\delta}_{\alpha'}) - 1  \right]e^{i\mathbf{q} \cdot \boldsymbol{\delta}_{\alpha'}}\left(\frac{\mathbf{k} \cdot \boldsymbol{\delta}_{\alpha'}}{||\mathbf{k}||}\right)(\boldsymbol{\delta}_{\alpha'} \cdot \boldsymbol{\epsilon}_{k}).
    \end{split}{}
\end{equation}{}
The magnon-phonon hybridization term gives the following contribution:
\begin{equation}\label{D2}
    \begin{split}
        \Sigma^{3}(k,i\nu_n) &=2\boldsymbol{\gamma}_k^\dagger \bar{G}_0^s(k, i\nu_n)\boldsymbol{\gamma}_k.
    \end{split}{}
\end{equation}{}
The change of sound velocity is then evaluated in the limit $k\rightarrow 0$. In this limit the $\gamma_k^1$ term is negligible compared to the $\gamma_k^2$ term and $E_k^\pm = E_{-k}^\pm$. We obtain an analytical expression for $\bar{R}_k$ by inverting Eq. (A4), and the self-energy is given by:
\begin{equation}\label{D3}
    \begin{split}
        \Sigma^{3}(k,i\nu_n) &=-\frac{\xi^2 S^{3}\cos\theta^2(\gamma_k^2)^2}{2\omega_k^0} \begin{pmatrix} 1 & 1 & -1 & -1 \end{pmatrix} \bar{R}_k \bar{G}_0^s(k, i\nu_n)\bar{R}_k^\dagger \begin{pmatrix} 1 \\ 1 \\ -1 \\ -1 \end{pmatrix} =-\frac{2\xi^2 S^{4}\cos\theta^2(\gamma_k^2)^2}{\omega_k^0}\frac{A_k+B_k+C_k+D_k}{(E_k^+)^2 - (i\nu_n)^2}.
    \end{split}{}
\end{equation}{}
For $v \gg v_{mag}$, we can neglect the $(E_k^+)^2 = v_{mag}^2 k^2 $ term in front of  $(i\nu_n)^2 = v^2 k^2$ in the denominator, and we finally obtain
\begin{equation}\label{D4}
    \begin{split}
   &\lim_{k\rightarrow 0} \frac{\Sigma^{3}(k,i\nu_n)_{i\nu_n \rightarrow \omega_k^0 + i0^+}}{\omega_k^0} = -\frac{\xi^2 |J_1|^3S^{4}}{v^4} \times \frac{1}{2} \left[\sum_{\delta_\alpha}p_{\delta_\alpha}  (\boldsymbol{\delta}_\alpha \cdot \boldsymbol{\epsilon}_{k})\left(\cos(\mathbf{q}_0 \cdot \boldsymbol{\delta}_\alpha) - 1 \right) \left(\frac{\mathbf{k} \cdot \boldsymbol{\delta}_\alpha}{||\mathbf{k}||}\right)\right]^2\\
   & \ \ \ \ \ \ \ \ \  \ \ \ \ \ \ \ \ \  \ \ \ \ \ \ \ \ \  \ \ \ \ \ \ \ \ \  \ \ \ \ \ \ \ \ \  \ \ \ \ \ \ \ \ \ \ \ \times\left[\sum_{\delta_\alpha}p_{\delta_\alpha}  \cos(\mathbf{q}_0 \cdot \boldsymbol{\delta}_\alpha)\left(\frac{\mathbf{k} \cdot \boldsymbol{\delta}_\alpha}{||\mathbf{k}||}\right)^2\right]\cos\theta^2\sin\theta^2.
    \end{split}{}
\end{equation}{}
This contribution has an extra factor $|J_1|^2S^2/v^2$ and is thus negligible in front of $\Sigma_1$ and $\Sigma_2$.

\

The magnon-bubble contribution is given by:
\begin{equation}\label{D9}
    \begin{split}
        \Sigma^4(k,i\nu_n) &= \sum_{q,i\nu_m,i}e^{-i\nu_m 0^-}\left(\bar{G}_0^s(q,i\nu_m)\right)_{ii}\left(\bar{N}_{k,-k-q,q}\right)_{ii} +\sum_q \Delta A_{qk}.
    \end{split}{}
\end{equation}{}
In the $k \rightarrow 0$, $k' \rightarrow -k$ limit, we replace $\left(1-e^{i\mathbf{k} \cdot \boldsymbol{\delta}_\alpha}\right)$ by $-i\left(\mathbf{k} \cdot \boldsymbol{\delta}_\alpha\right)$ and by taking out the prefactors, we obtain:
\begin{equation}\label{D10}
    \begin{split}
        &\lim_{k\rightarrow 0} \frac{\Sigma^4(k,i\nu_n)_{i\nu_n \rightarrow \omega_k^0 + i0^+}}{\omega_k^0} = \frac{\xi^2 S |J_1|}{v^2} \times \left(\frac{1}{4N} \sum_q \left[ \bar{N}_q^{22} + \bar{N}_q^{44}\right] + \Sigma_1\right),
    \end{split}{}
\end{equation}{}
where the matrix $\bar{N}_{q}$ is defined by
\begin{equation}\label{D11}
    \begin{split}
        \bar{N}_{q} &= \left(\bar{R}_{q}\right)^\dagger \begin{pmatrix} A_{q}'' & C_{q}'' & B_{q}'' & D_{q}'' \\ (C_{q}'')^* & A_{q}'' & (D_{q}'')^* & (B_{q}'')^* \\ B_{q}'' & D_{q}'' & A_{-q}'' & (C_{-q}'')^* \\ (D_{q}'')^* & (B_{q}'')^* & C_{-q}'' & A_{-q}'' \end{pmatrix} \bar{R}_{q} \\
    A_{q}'' &= \sum_{\delta_\alpha}p_{\delta_\alpha}\left[\cos\theta^2\cos(\mathbf{q}_0 \cdot \boldsymbol{\delta}_\alpha) + \sin\theta^2\right]\left[\left(\frac{\mathbf{k} \cdot \boldsymbol{\delta}_\alpha}{||\mathbf{k}||}\right)(\boldsymbol{\delta}_\alpha \cdot \boldsymbol{\epsilon}_{k})\right]^2\\
    &- \frac{p_2}{2} \sum_{\delta_2} \left[ \left(1 + \sin\theta^2 \right)\cos(\mathbf{q}_0 \cdot \boldsymbol{\delta}_2) + \cos\theta^2 +2i\sin\theta\sin(\mathbf{q}_0 \cdot \boldsymbol{\delta}_{2}) \right] e^{i\mathbf{q} \cdot \boldsymbol{\delta}_2}\left[\left(\frac{\mathbf{k} \cdot \boldsymbol{\delta}_2}{||\mathbf{k}||}\right)(\boldsymbol{\delta}_2 \cdot \boldsymbol{\epsilon}_{k})\right]^2 \\
    B_{q}'' &= -\frac{p_2}{2} \sum_{\delta_2} \cos\theta^2\left[\cos(\mathbf{q}_0 \cdot \boldsymbol{\delta}_2) - 1  \right]e^{i\mathbf{q} \cdot \boldsymbol{\delta}_2}\left(\frac{\mathbf{k} \cdot \boldsymbol{\delta}_2}{||\mathbf{k}||}\right)(\boldsymbol{\delta}_2 \cdot \boldsymbol{\epsilon}_{k})\\
    C_{q}'' &= -\frac{1}{2}\sum_{\delta_{\alpha'} = \delta_1, \delta_3} p_{\alpha'} \left[\left(1 + \sin\theta^2 \right)\cos(\mathbf{q}_0 \cdot \boldsymbol{\delta}_{\alpha'})+\cos\theta^2 +2i\sin\theta\sin(\mathbf{q}_0 \cdot \boldsymbol{\delta}_{\alpha'}) \right]e^{i\mathbf{q} \cdot \boldsymbol{\delta}_{\alpha'}}\left[\left(\frac{\mathbf{k} \cdot \boldsymbol{\delta}_{\alpha'}}{||\mathbf{k}||}\right)(\boldsymbol{\delta}_{\alpha'} \cdot \boldsymbol{\epsilon}_{k})\right]^2\\
    D_{q}'' &= -\frac{1}{2}\sum_{\delta_{\alpha'} = \delta_1, \delta_3} p_{\alpha'}   \cos\theta^2\left[\cos(\mathbf{q}_0 \cdot \boldsymbol{\delta}_{\alpha'}) - 1  \right]e^{i\mathbf{q} \cdot \boldsymbol{\delta}_{\alpha'}}\left[\left(\frac{\mathbf{k} \cdot \boldsymbol{\delta}_{\alpha'}}{||\mathbf{k}||}\right)(\boldsymbol{\delta}_{\alpha'} \cdot \boldsymbol{\epsilon}_{k})\right]^2.
    \end{split}{}
\end{equation}{}
The sum is evaluated using standard Monte Carlo integration over the first Brillouin zone of the diamond lattice. This contribution is negligible in front of $\Sigma_1$ and $\Sigma_2$.
\section{Magnetization anisotropy}
\begin{figure}[h!]
    \centering
    \includegraphics[width=\linewidth]{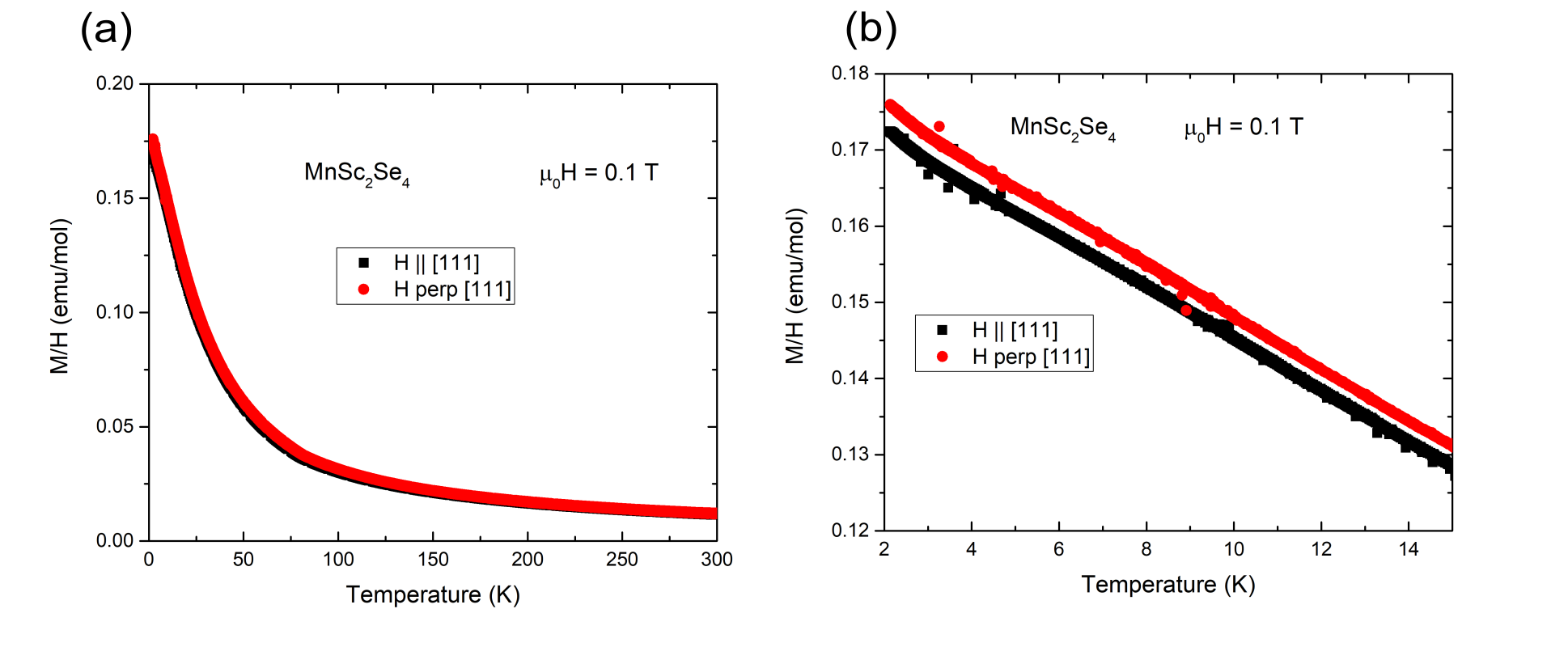}
    \caption{(a) Temperature-dependent magnetization of single-crystalline MnSc$_2$Se$_4$ for two different field directions. (b) Enlarged view at low temperatures.}\label{figS1}
\end{figure}

As shown in Fig. \hyperref[figS1]{\ref*{figS1}}, a small anisotropy (2\%) of the magnetization in MnSc$_2$Se$_4$ is observed.
\bibliography{Biblio/biblio_JSourd_MNSc2Se42.bib}